\documentclass[12pt,a4paper]{article}
\pdfoutput=1
\usepackage{color}
\newcommand{\comment}[1]{}
\usepackage[ruled,vlined,linesnumbered,noend]{algorithm2e}

\usepackage{amsthm,color,graphicx}
\usepackage[utf8]{inputenc}
\usepackage{amsmath}
\usepackage{amsfonts}
\usepackage{amssymb}
\usepackage{bigstrut}
\usepackage{booktabs}
\usepackage{placeins}
\usepackage{makeidx}
\usepackage{multirow}
\usepackage{subfloat}
\usepackage{graphicx}
\usepackage{lmodern}
\usepackage{lipsum}
\usepackage{titlesec}
\usepackage{setspace}
\usepackage[hidelinks]{hyperref}

\usepackage{caption}
\usepackage{subcaption}

\titleformat{\section}
{\normalfont\fontsize{14}{15}\bfseries}{\thesection}{1em}{}
\usepackage{mathptmx}

\usepackage[left=1in,right=1in,top=1in,bottom=1in]{geometry}

\date{}
\begin{document}

\SetKw{KwAnd}{and}
\SetKwFunction{GetPotentialCandidates}{GetPotentialCandidates}
\SetKwFunction{suff}{suff} 
\SetKwData{TRUE}{TRUE}
\SetKwData{FALSE}{FALSE}
\SetKwData{CS}{CurSuff}
\SetKwData{CSt}{CurSt}
\SetKwData{fs}{FutStep} 
\SetKwData{TT}{MTT}
\SetKwData{IW}{win}
\SetKwData{PC}{ParCor}

\SetKwData{TSPI}{TimeStepAheadInd}
\SetKwData{ETB}{ExpTimeBin}
\SetKwData{CTB}{CurrTimeBin}
\SetKwData{Fl}{Flag}
\newcommand{\lA}{\ensuremath{\leftarrow}}
\newcommand{\scr}[1]{\ensuremath{\mathcal{#1}}}

\newcommand{\B}[1]{\ensuremath{\mathbf{#1}}}
\newtheorem{proposition}{Proposition}
\newtheorem{property}{Property}
\newtheorem{definition}{Definition}

\setcounter{page}{1}

{\begin{center}
\textbf{{\Large Bus Travel Time Prediction: A log-normal Auto-Regressive (AR) Modeling Approach}}
\end{center}}
\vspace{0in}

\begin{center}
\textbf{B. Dhivyabharathi\footnote{Doctoral Student, Department of Civil Engineering, Indian Institute of Technology Madras, Chennai, INDIA 600036, E-mail: gebharathi@gmail.com}
B. Anil Kumar\footnote{Senior Project Officer, Department of Civil Engineering, Indian Institute of Technology Madras, Chennai, INDIA 600036, E-mail: raghava547@gmail.com}, Avinash Achar\footnote{Research Scientist, TCS Innovation Labs, Chennai, INDIA 600113, E-mail: achar.avinash@tcs.com}, Lelitha Vanajakshi\footnote{*Corresponding Author, Professor, Department of Civil Engineering, Indian Institute of Technology Madras, Chennai, INDIA 600036, E-mail: lelitha@iitm.ac.in}*}
\end{center}
\vspace{0.2in}

\noindent \textbf{ABSTRACT}\\
\noindent Providing real-time arrival time information of the transit buses has become inevitable in urban areas to improve the efficiency of the public transportation system. However, accurate prediction of arrival time of buses is still a challenging problem in dynamically varying traffic conditions especially under heterogeneous traffic condition without lane discipline. One broad approach researchers have adopted over the years is to divide the entire bus route into sections and model the correlations of section travel times either spatially or temporally. The proposed study adopts this approach of working with section travel times and developed two predictive modelling methodologies namely (a) classical time-series approach employing a seasonal AR model with possible integrating non-stationary effects and (b) linear non-stationary AR approach, a novel technique to exploit the notion of partial correlation for learning from data to exploit the temporal correlations in the bus travel time data. Many of the reported studies did not explore the distribution of travel time data and incorporated their effects into the modelling process while implementing time series approach. The present study conducted a detailed analysis of the marginal distributions of the data from Indian conditions (that we use for testing in this paper). This revealed a predominantly log-normal behaviour which was incorporated into the above proposed predictive models. Towards a complete solution, the study also proposes a multi-section ahead travel time prediction algorithm based on the above proposed classes of temporal models learnt at each section to facilitate real time implementation. Finally, the predicted travel time values were corroborated with the actual travel time values. From the results, it was found that the proposed method was able to perform better than historical average, exponential smoothing, ARIMA, and ANN methods and the methods that considered either temporal or spatial variations alone. 

\vspace{0.15in}


\section{INTRODUCTION}
\noindent Real-time bus arrival information is highly sought-after information in recent years to enhance the efficiency and competitiveness of public transport \cite{ceder:07}. However, uncertainties in arrival time are quite common in public transit due to dynamic traffic conditions, particularly for highly varying heterogeneous traffic condition, contributed by various factors such as lack of lane discipline, fluctuating travel demand, incidents, signal timing and delay, bus stop dwell times, seasonal and cyclic variations, etc. In countries such as India in particular, public transport buses scarcely stick to any predefined schedule while bus commuters hardly have any real-time information of the likely arrival time of the bus they are expecting. Given the lack of this crucial real-time information, commuters may end up taking private vehicles to reach their respective destinations. This may lead to reduction in modal share of public transport and increase in composition of private vehicles contributing towards the raise in congestion and other related negative impacts. Hence, prediction of travel/arrival time and informing the same to passengers is inevitable to make public transport more attractive, efficient, and competitive especially in urban areas. Such real-time information can also be used to assist commuters in making better trip-related decisions ahead of the journey, which significantly reduces anxiety levels while waiting for a bus \cite{ watkins:11,brake:14,cats:15}. Thus, it is evident that the negative impacts associated with lack of reliability of the existing public transportation system can be reduced to an extent by predicting accurate arrival time of buses and communicating it to the passengers in advance. 

The travel time prediction methodologies can be broadly grouped into data driven and traffic flow theory based methods. With the advent and implementation of diverse modern sensing
technologies that generate large amount of data, data-driven techniques are getting more popularity. For example, these days, many urban public transportation systems deploy
Automated Vehicle Location (AVL) systems like Global Positioning System (GPS) to monitor the position of buses in real time, which can provide a constantly growing database of location and timing details
\cite{gent:18}. Such data can rightly be used for developing a Bus Arrival Time Prediction (BATP) system , as the collected data inherently capture the nature and patterns of
traffic in real time. Many researchers have explored the modelling and prediction of bus travel time, but only a few studies have considered exploiting the temporal correlations in the
AVL data systematically while also taking into consideration the statistical distributions followed by the
travel time data. This work tries to bridge this gap in existing literature appropriately using linear statistical models. 

The dynamic bus travel time prediction problem considered in this paper is to accurately predict the travel times at sections ahead from the current position of the bus 
as the trip is in progress in real time. To achieve this, exploiting the
information of the travel times experienced by the previous buses in the subsequent sections is a promising choice. This means utilizing the temporal dependencies in the
spatio-temporal AVL data of all the buses. Hence, the travel times experienced at different time intervals of the day at a particular section were chosen as observations of our
time series or univariate sequential data. A statistical sequential predictive model using a novel linear, non-stationary approach that performs learning exploiting
the notion of partial correlation is developed.  This study also proposes a time-series approach employing a seasonal AR model with possible integrating non-stationary effects for
the prediction of bus travel time. Both the temporal predictive modelling approaches predict ahead in time at each segment independently. To appropriately and correctly utilize these temporal predictions available at each section, the present study additionally formulated a multi sections ahead prediction framework to facilitate real time implementation of the developed BATP system.

The present study also carried out a detailed analysis of the marginal distributions of the travel time data collected from typical Indian conditions (that we use for testing in
this paper) which revealed a predominantly log-normal behaviour.  
 The lognormal nature of the data is incorporated into both the above predictive models and statistically optimal prediction schemes in the lognormal sense are utilized for all
predictions. Our experiments clearly revealed superior predictions
in favor of prediction models which explicitly utilized the lognormal nature of the data.

\section{Literature Review}

\noindent Various studies have been reported on prediction of travel times and the methods used can be broadly classified into traffic flow-theory based and data-driven methods.
\FloatBarrier

Traffic flow theory based methods establish a mathematical relationship between appropriate variables that tries to capture the system characteristics. Such methods develop models based on traffic flow theory, which is mostly based on the first principles of physics. Such theory-based approaches usually focus on recreating the traffic conditions in the future time intervals and then deriving travel times from the predicted traffic state \cite{van:04}. Such theory based models can be classified as a) Macroscopic, b) Microscopic, and c) Mesoscopic, based on the level of detail it captures \cite{mihay:04, hins:07, miska:05, kumar:17}. Majority of the studies under this category used the conservation of vehicle principle to predict travel time under homogenous traffic condition \cite{nam:96, van:04} and  a few studies developed simulation approaches \cite{celi:13}, dynamic traffic assignment based methods \cite{long:11} for the same. However, all of the above studies predicted travel times from flow/speed measurements under homogenous traffic conditions. It is very standard to use a suitable estimation/prediction tool to predict the traffic state variables recursively in real time, when model based methods are implemented. Most of the studies for travel time prediction used Kalman Filtering Technique (KFT) as the estimation tool \cite{wall:99,nantha:03,chu:05,leli:09, padma:09,kumar:17a}. A few studies have explored some non-linear estimation tools such as Particle Filtering, Extended Kalman Filtering, Unscented Kalman Filtering, etc. for the purpose of travel time prediction \cite{mihay:04, dhivya:18}. Due to the non-linearity, complexity, and uncertainty of contributing factors of the traffic conditions, traffic flow theory based methods needs highly complex and sophisticated models and recursive techniques to capture the system dynamics under high variation condition.

Data driven approaches use larger databases to develop statistical/empirical relations to predict the future travel time without really representing the physical behavior of the modelled system \cite{zhang:17}. In a way, these methods exploit the available data to extract the system characteristics. Various data driven approaches reported in the literature include historical averaging methods, empirical methods, statistical methods, and machine learning methods. 

Historic averaging methods \cite{lin:99,kuchi:03} predict the current and future travel time by averaging the historical travel time of previous journeys. These methods assume that traffic patterns are cyclical and the projection ratio (the ratio of the historical travel time in a particular link to the current travel time) will remain constant. Hence, historic averaging methods are not suitable for highly varying traffic conditions, where travel times experience large variations.

Empirical models are based on intuitive empirical relation or mathematical concepts. A few studies \cite{van:02,shalaby:04,boel:06,fei:11,hage:12,chen:14,celena:17} have explored and showed the superiority of the empirical model combined with recursive filtering over the other methodologies. However, the efficiency of the method is highly dependent on the presumed relationship or assumptions. 

Machine learning techniques such as Artificial Neural Network (ANN) and Support Vector Machine (SVM) are some of the most commonly reported prediction techniques for travel time prediction because of their ability to solve complex relationships \cite {wu:03, chien:02}. A few studies \cite{van:02, chen:04, liu:06, van:06} combined the machine learning techniques such as SVM, ANN with filtering algorithms such as Kalman Filtering to predict travel time. \cite{bin:06} and \cite{leli:04} have compared the performance of ANN and SVM and reported that SVM outperforms ANN in the prediction of arrival times and also it was reported that SVM is not susceptible to the over fitting problem unlike traditional ANN, as it could be trained through a linear optimization process. Many studies \cite{chen:04, bin:06, yu:11,julio:16,yin:17} have reported better performance of machine learning techniques compared to other existing methods. However, machine learning techniques require a large amount of data for training and calibration, and are computationally expensive, restricting their usage for wide real time applications. 

Statistical methods such as regression and time series  predicts the future values by developing relationship among the variables affecting travel time or from the series of data points listed in time order. A few studies have implemented regression methods \cite{abdel:98,patna:04,schr:04,kwon:05,rk:06,kwon:07,log:08,Yu:17,tira:17,zhou:17} for arrival time prediction. \cite{yu:09} integrated regression methods with adaptive algorithms to make it suitable for real time implementations. However, the accuracy of the prediction using this approach highly depends on identifying and applying suitable independent variables. This requirement limits the applicability of the regression model to the transportation areas as variables in transportation systems are highly inter-correlated \cite{bae:95}. However, the effect of multicollinearity can be reduced by two approaches – by calculating the value of Variance Inflation Factor (VIF), or use robust regression analysis instead of ordinary least squares regression, such as ridge regression \cite{haworth:14}, lasso regression \cite{hammer:10}, and principal component regression \cite{park:81}. Statistical learning regression methods such as regression tree \cite{vena:02}, bagging regression \cite{wolf:15}, and random forest regression \cite{hammer:10} are also used. 

Time series models (or data-driven approaches in general) are based on the assumption that the current and future patterns in data largely follow patterns similar to 
historically observed data. \cite{chen:12} stated that this
method greatly depends on historic data to develop relations for forecasting future time variations. Travel time prediction domain has seen numerous time series works with
various modelling strategies and techniques. Earlier studies \cite{ oda:90,arem:97,lee:98, angelo:99, guin:06} estimated travel time from other traffic flow variables such as
traffic volume, speed, and occupancy collected from location-based sensors such as loop detectors and implementations reported were mainly on freeways. 
 
Literature has also seen time series appraoches for travel time prediction by modelling
several other related variables such as delays \cite{bhandari:05}, headways \cite{andres:17}, dwell time \cite{rashidi:15}, running speed \cite{xinghao:13}, etc. Only a few studies modelled the travel time observations directly for a BATP problem. In addition, most of those studies considered either entire trip travel time \cite{jairam:18} or bus stop to bus stop travel time \cite{bhandari:05}, which eventually may not capture the variations in travel time, if the considered stretch is very long. The present study adopted an effective way of segmenting the entire route into smaller sections of uniform length.  The time-series observations in the proposed method are different from all these existing time series approaches for analysing bus travel time data. {\em At each section, the travel times experienced at different times of the day are modelled as a separate time series.} Hence, the proposed approach can potentially capture the variations in travel time better than earlier methods.  

In a traffic system, the travel times also exhibit  patterns which are mostly cyclical with variations in magnitude depending on the  spatial and geometric features of the location. 
For example, presence of a signal at a particular location mostly results in higher travel times and therefore, this trend/pattern would clearly reflect in
the historical travel time data. Hence, the present study adopted a time series (or sequential modelling) framework on travel time observations at different times of the day 
for bus arrival time prediction. For each section, the proposed methodology models the
temporal correlations between travel times experienced at different times of the day. 
These temporal correlations are learnt using two approaches: (i) based on a novel application of the notion of partial correlation and (ii) using a seasonal AR time-series approach.
 
\vspace{12pt}

\noindent\textbf{Studies under Heterogeneous Traffic Conditions:} A few methodologies were developed to predict bus travel time under heterogeneous lane-less traffic condition and
are discussed here \cite{rk:06, leli:09, padma:09}. Majority of the above studies used a space discretization approach to predict bus arrival times \cite{leli:09, kumar:13}. Here,
the route was spatially discretized into smaller sections and hypothesized a relation in travel time between neighbouring sections, i.e., the travel time of a bus in a particular
section can be predicted using the travel time of the same bus in the previous section. This was adopted mainly due to the lack of availability of a good historic data base.
\cite{padma:09} extended the work by separating dwell times and running time from total travel time and modelled their characteristics independently. They found that considering
dwell times does not bring a significant improvement in overall prediction accuracy. However, such a method may not be able to capture  correlations meaningfully in all scenarios. For example, the travel times in neighbouring sections may not be dependent on each other if one had a signal and the other is not influenced by that signal. In such cases, prior travel time information  from the same section may be a better input for prediction than the neighbouring section’s travel time. \cite{kumar:17a} hypothesized a relation based on temporal evolution of travel time and \cite{kumar:17} developed a model based approach to explore the spatio-temporal evolutions in travel time. Irrespctive of the modelling nature, the aforementioned studies neither analysed nor incorporated the distributional aspects of travel time into predictive modelling. The current study incorporates this feature into the predictive model by exploiting the predominantly lognormal nature of the data  before exploring two sequential modelling approaches to model travel time evolutions across the day to predict bus arrival times. 

This paper is organized as follows. First, the details of study site and data processing are presented. Next, the section on data analysis discusses the various analysis conducted to assess the characteristics and variations of travel time. This is followed by methodology section that explains the travel time prediction schemes developed to incorporate the identified characteristics of the data using the concepts of time series analysis. Results section is presented next discussing the evaluation of the proposed approaches using real-world data. Finally, the study is concluded by summarizing the findings.
 
\section{DATA COLLECTION}
\begin{figure}[htbp]
	\centering
	\includegraphics[width=0.35\linewidth]{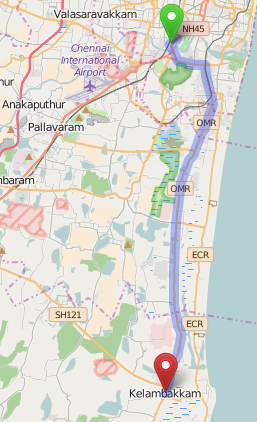}
	\caption{Study stretch considered (19B bus route).}
	\label{fig:route}
\end{figure}

\noindent Probe vehicles fitted with GPS units are commonly used to collect data for advanced public transportation system applications. In the present study, data were collected using permanently fixed GPS units in Metropolitan Transport Corporation (MTC) buses in the city of Chennai, India. In this study, MTC
bus route 19B was considered, which has a route length of around 30 km as shown in Figure \ref{fig:route}. The selected route has 25 bus stops and 14 signalized intersections, many of which are severely congested and
oversaturated. The study stretch represents typical heterogeneous, lane-less Indian traffic and includes varying geometric characteristics, volume levels, and land use characteristics. The selected route covers both urban and rural areas and has both four-lane and six-lane roads. The Average Daily Traffic (ADT) at one location with a counting station in the selected route was measured to be around 40,000 vehicles. Because of lack of availability of any other counting stations along the route, this number can be considered as a representative value. The vehicle composition in this stretch is reported to be 47\% two-wheelers, 7.3\% three-wheelers, 43.7\% light motor vehicles (LMVs), and 2\% heavy motor vehicles (HMVs) \cite{asha:14}. In the test bed identified, there are no exclusive lanes for the public transport buses. They share the road space with all other vehicle modes. The average headway between successive buses is around 20 minutes for the selected route. While overtaking of buses is feasible, overtaking of a bus by the next bus of the same bus route is very rare.

In this study, 34 days of GPS data collected at every 5 seconds was used. Out of these 34 days, the last one week’s data were kept for corroboration
purpose while   the remaining 27 days' data were use for model building. The collected GPS data included the ID of the GPS unit, time stamp, and the location details of
the bus in terms of latitude and longitude at every 5 seconds. The raw data obtained from GPS were then sorted and checked for quality. This included trip identification,
data processing and data cleaning. Trip identification is the process through which multiple trips made by a bus in a particular day were separated. This process also
involved the separation of onward and return journeys of the same trip. Data processing involved conversion of the raw location details (latitude and longitude) into
distance, using Haversine formula \cite{reid:14}. The difference between time stamps of two consecutive GPS data points gives the travel time for the distance between the
selected two points. After this process, the data consisted of travel times and the corresponding distance between consecutive locations for all the trips. In this study,
the entire route was divided into smaller sections of 500 m length as it is the average distance found between two successive bus stops and analyses were carried out. Data cleaning was carried out to remove outliers from the data. The
outliers were identified by assuming free flow speed of 60 km/h as the maximum speed leading to the lower bound of travel time as 30 s for a 500 m section. The upper bound was kept as 95\textsuperscript{th} percentile travel time for sections. Travel times less than lower bound and more than upper bound were replaced with its bound permissible values. After all these steps, the final data set consisted of the date, time, and 500 m section travel times for every trip.

\section{PRELIMINARY DATA ANALYSIS}
 \noindent In this study, at each section, the travel times observed at different times of the day are  viewed as sequential data or a time series. The entire 24 hr time window is divided into one hour slots. Due to absence of data for
	about $5$ hours in the night, we have $19$ active slots per  day. The data is pre-processed such that at each time slot there is only one travel time observation.
It was assumed that there are correlations between travel times of adjacent time slots and our proposed predictive modelling strategies to be explained in the next section precisely 
capture this.

\begin{figure}[htbp]
	\centering
	\begin{subfigure}{0.4\textwidth}
		\centering
		\includegraphics[width=\textwidth]{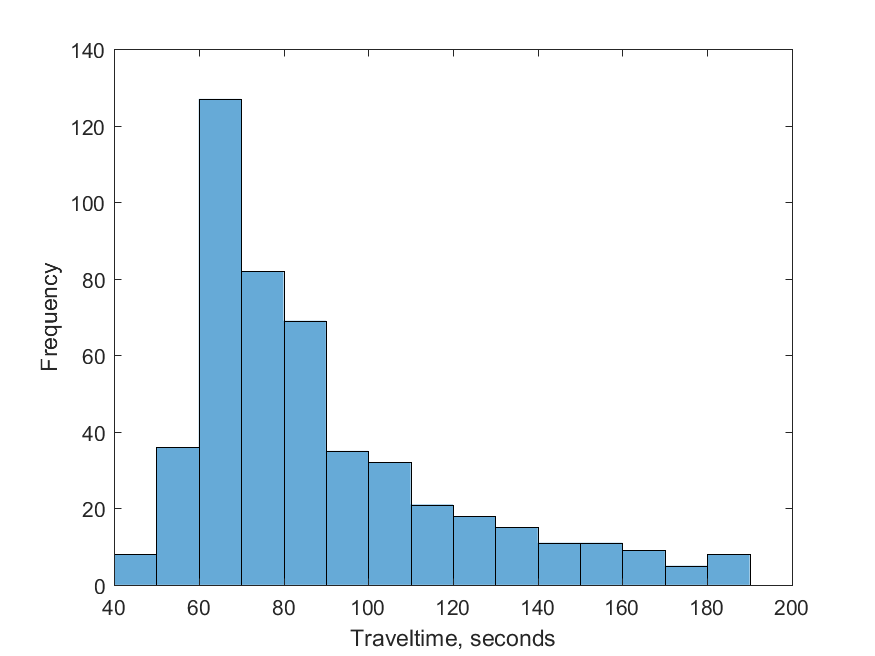}
		\caption{Section 13}
		\label{fig:7}
	\end{subfigure}	
	\begin{subfigure}{0.4\textwidth}
		\centering
		\includegraphics[width=\textwidth]{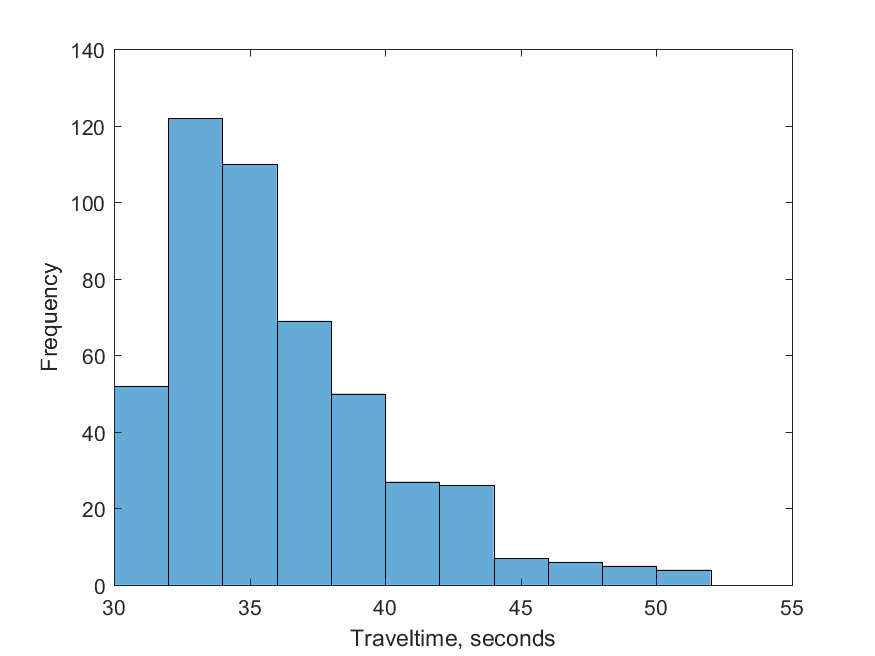}
		\caption{Section 20}
		\label{fig:27}
	\end{subfigure}
	\begin{subfigure}{0.4\textwidth}
		\centering
		\includegraphics[width=\textwidth]{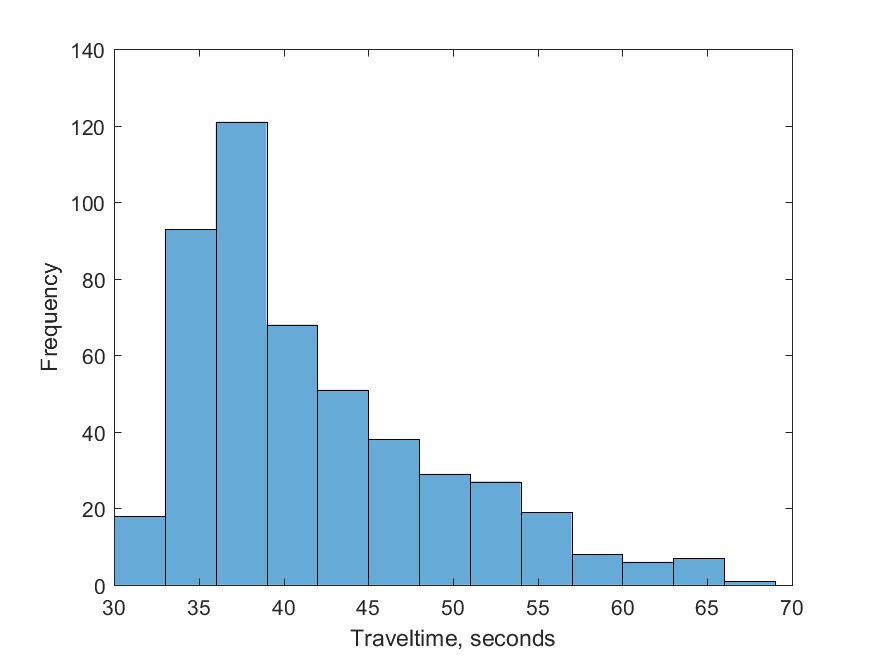}
		\caption{Section 31}
		\label{fig:40}
	\end{subfigure}	
	\begin{subfigure}{0.4\textwidth}
		\centering
		\includegraphics[width=\textwidth]{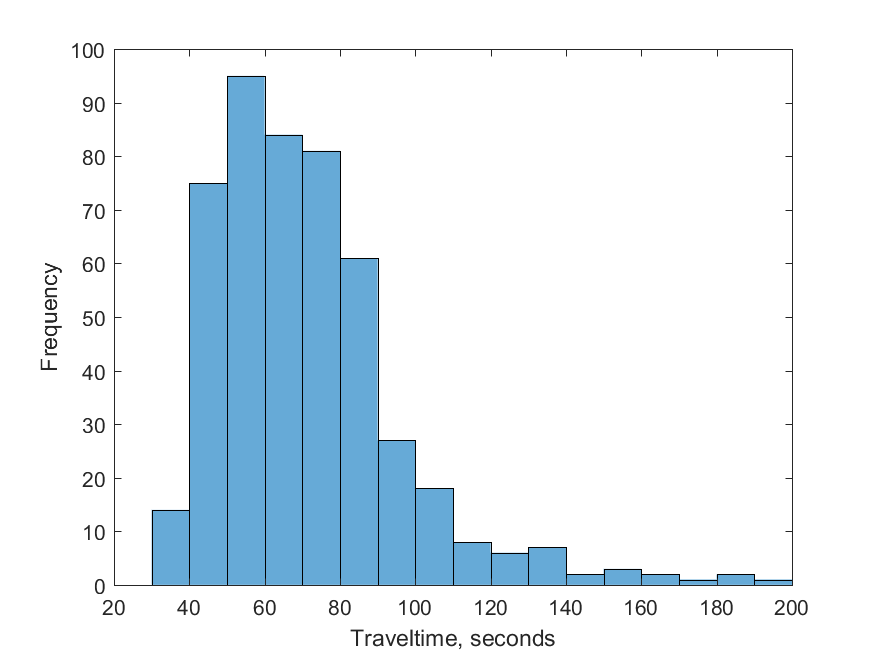}
		\caption{Section 45}
		\label{fig:45}
	\end{subfigure}	
	\caption{Histogram for sample sections.}
	\label{fig:histogram}	
\end{figure}

\begin{figure}[htbp]
	\centering
	\includegraphics[width=0.83\linewidth]{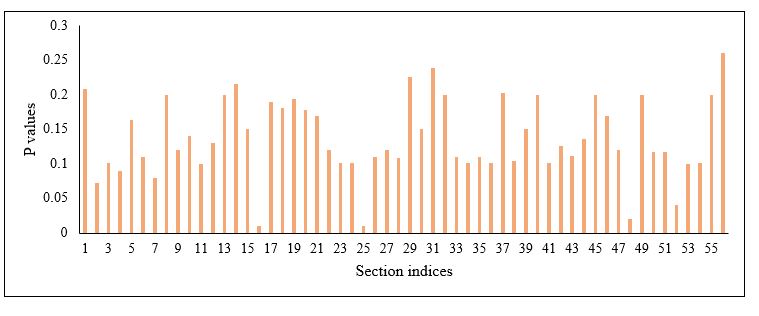}
	\caption{\textit{p}-values obtained from K-S test for each section across the route.}
	\label{fig:p_values}
\end{figure}

\begin{figure}[!h]
	\centering
	\begin{subfigure}[b]{0.49\textwidth}
		\centering
		\includegraphics[width=0.95\textwidth]{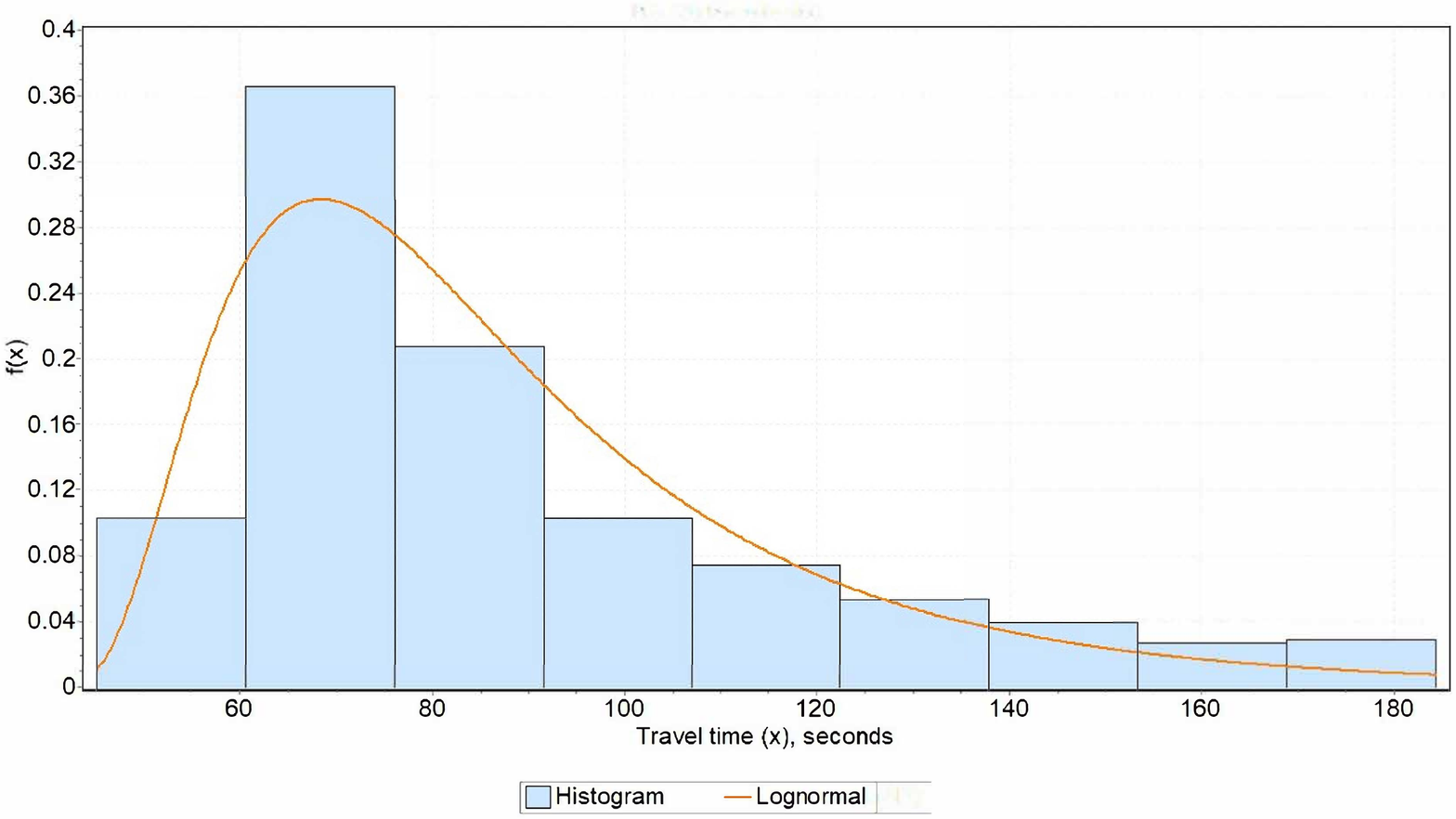}
		\caption{Section 13}
		\label{fig:easy7}
	\end{subfigure}	
	\begin{subfigure}[b]{0.49\textwidth}
		\centering
		\includegraphics[width=0.95\textwidth]{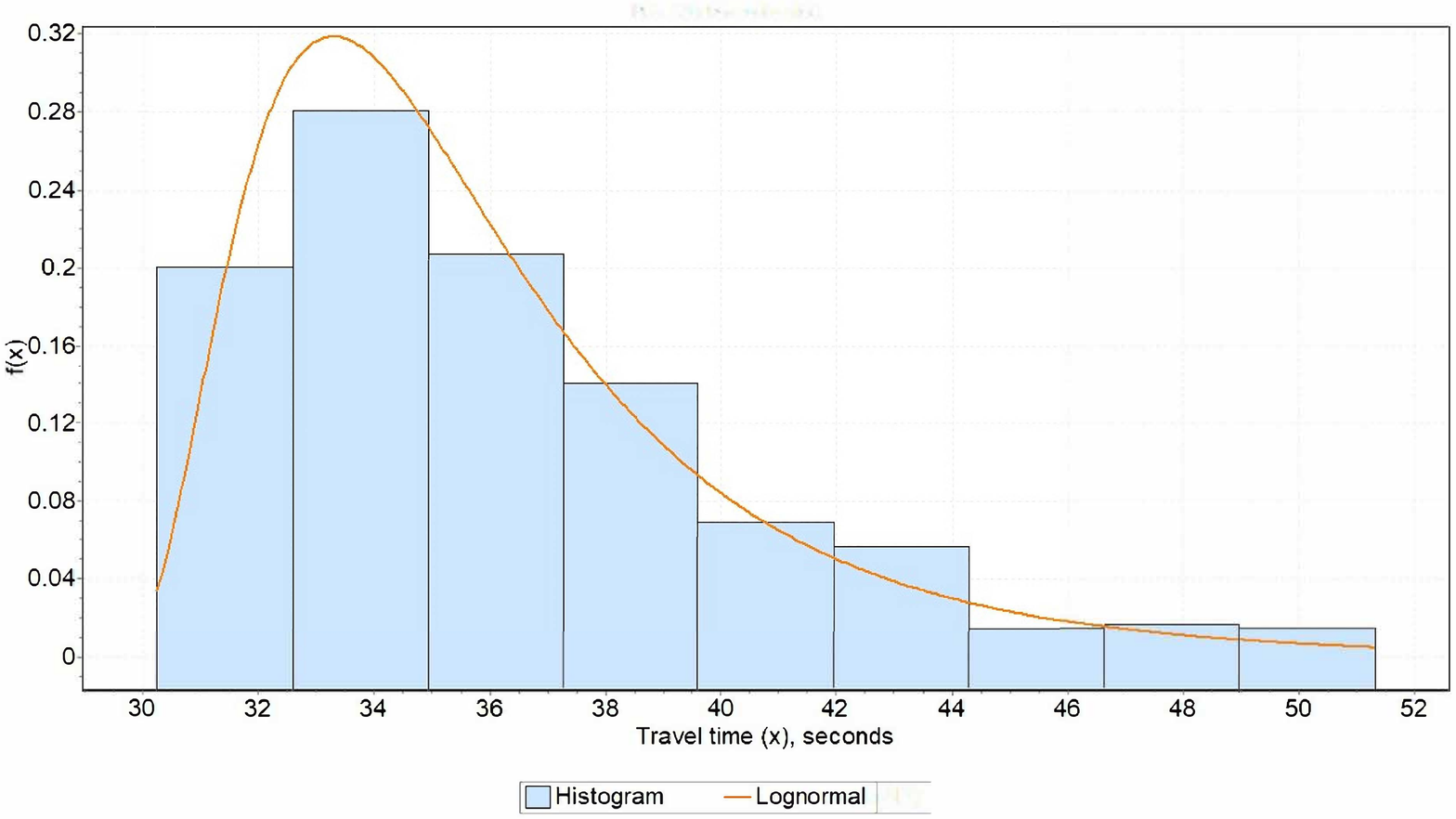}
		\caption{Section 20}
		\label{fig:easy27}
	\end{subfigure}
	\begin{subfigure}[b]{0.49\textwidth}
		\centering
		\includegraphics[width=0.95\textwidth]{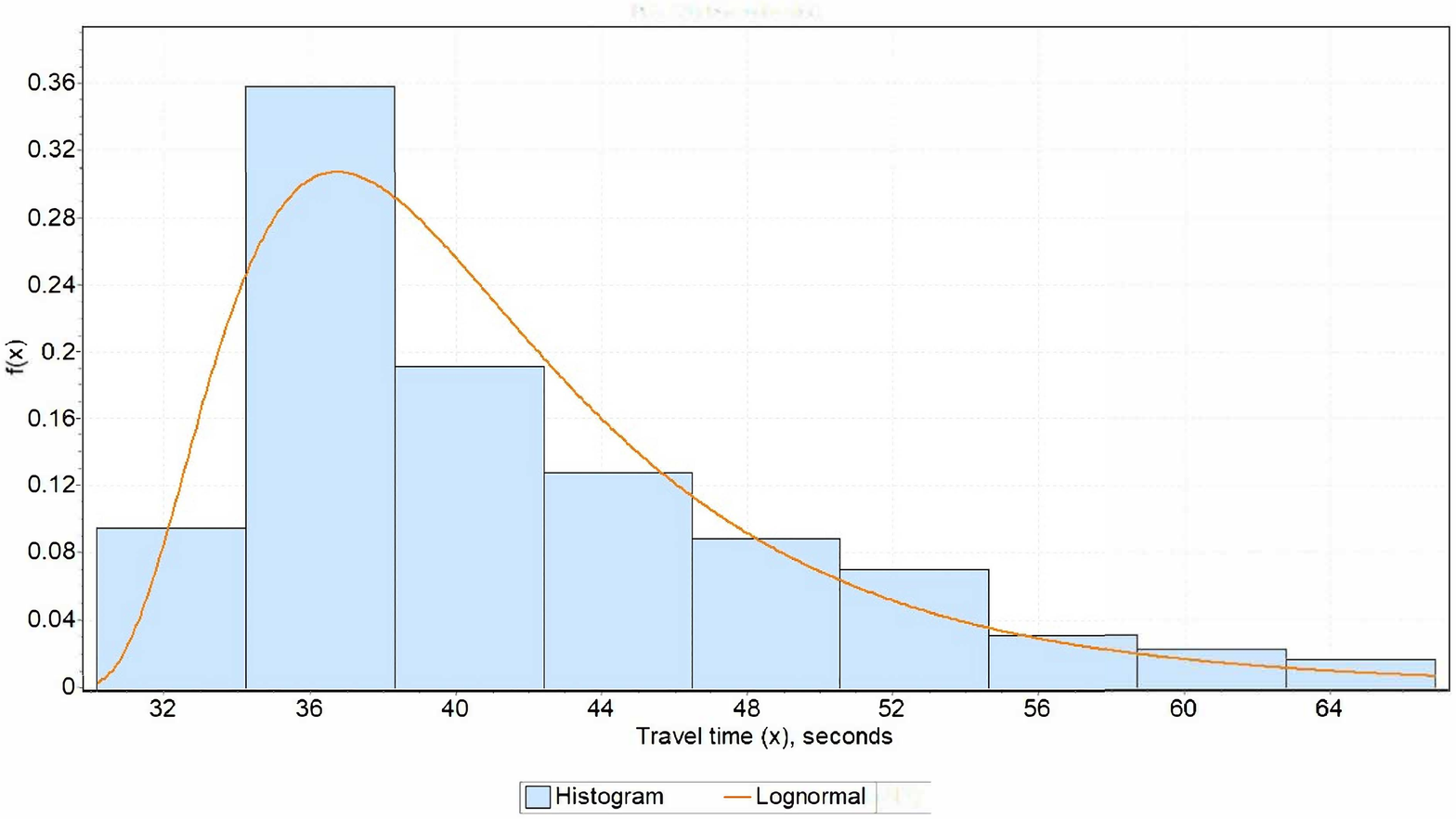}
		\caption{Section 31}
		\label{fig:easy40}
	\end{subfigure}	
	\begin{subfigure}[b]{0.49\textwidth}
		\centering
		\includegraphics[width=0.95\textwidth]{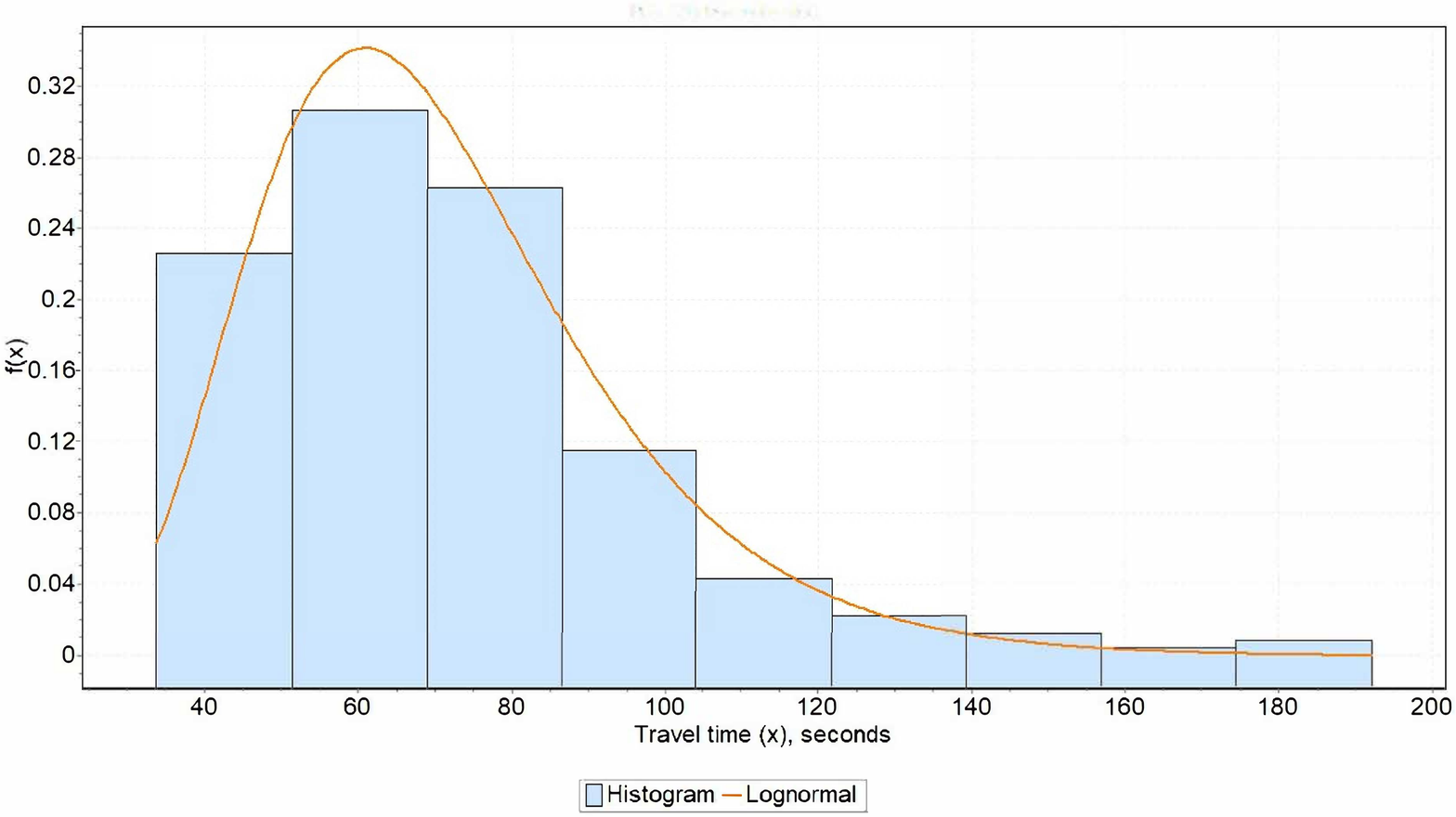}
		\caption{Section 45}
		\label{fig:easy45}
	\end{subfigure}	
	\caption{Distribution fitting results for sample sections.}
	\label{fig:easy}	
\end{figure}

Understanding the underlying data distribution is a key step before applying any statistical modelling approach \cite{rahman:18}. The processed data observations were grouped section-wise and distribution fitting was carried  out for each section. An analysis of travel time distribution was carried out  and a few sample histograms are shown in Figure ~\ref{fig:histogram}. From Figure~\ref{fig:histogram} it can be seen that the distribution is skewed to the right, which indicate a positively skewed distribution. The present study used the data distribution fitting software “Easyfit” to check the possible distributions to be tested. In the next level, codes were written in MATLAB to check the distribution fitting and the subsequent analysis confirmed the selection of lognormal distribution as the best fit distribution for the considered dataset. In order to check the log-normality of the data, a Kolmogorov-Smirnov test (K-S test) was conducted at 5\% significance level. The null hypothesis assumed for this test is that the data follows log-normal distribution. For majority of cases, the test failed to reject null hypothesis by showing \textit{p}-values greater than 0.05, indicating that the data sets can be approximated by log-normal distribution. Figure~\ref{fig:p_values} shows the results obtained from K-S test, i.e., \textit{p}-values for various sections along the route across all hours and days and  Figure~\ref{fig:easy} shows sample plots of the best log-normal fit for the respective empirical distributions of Figure~\ref{fig:histogram}.
\vspace{-0.1in}

\vspace{-0.2in}

\FloatBarrier

As the analysis results showed that the marginal distribution of data is mostly log-normal, the next stage was to model this data, incorporating its
log-normal characteristics.  The present study adopts concepts of time series to design prediction schemes under log normal data distribution assumption, 
which is described in the next section.

\section{METHODOLOGY}

\noindent  

As described in the previous section, a KS test showed lognormal distribution as the best fit among a set of standard distributions for modelling the marginal
distribution of the data. The pdf of a  univariate lognormal random variable $Y$ with parameters $\mu$ and $\sigma$ is defined as 
\begin{equation}
f_{Y}(y) = \frac{1}{x\sigma \sqrt{2\pi}} e^{-\frac{(ln(y) - \mu)^2}{2\sigma^2}}, 
\end{equation}
where $ln(Y)$ is a Gaussian random variable with mean $\mu$ and variance $\sigma^2$.
Accordingly, the travel time observations are assumed to follow a lognormal process. 
Prediction schemes, which exploit the lognormality exhibited by the data and make statistically optimal predictions are proposed.
\noindent This study explores temporal modelling and  prediction in primarily  two  ways. In the
first approach (Seasonal AR modelling), the non-stationarity of the data is tackled by removing possible deterministic trends and seasonality in the data before applying standard stationary model fits. In the
second approach (Non-stationary method), a general non-stationary data-driven model with a principled method to  estimate the time-varying parameters of the non-stationary model is proposed.    
In these proposed approaches, both the Gaussian and log-normal assumptions on the data are explored and are disscussed below.

\subsection{Seasonal AR Modelling with Possible Integrating Effects}
\begin{figure}[htbp]
	\centering
	\begin{subfigure}{0.45\textwidth}
		\includegraphics[width=\linewidth]{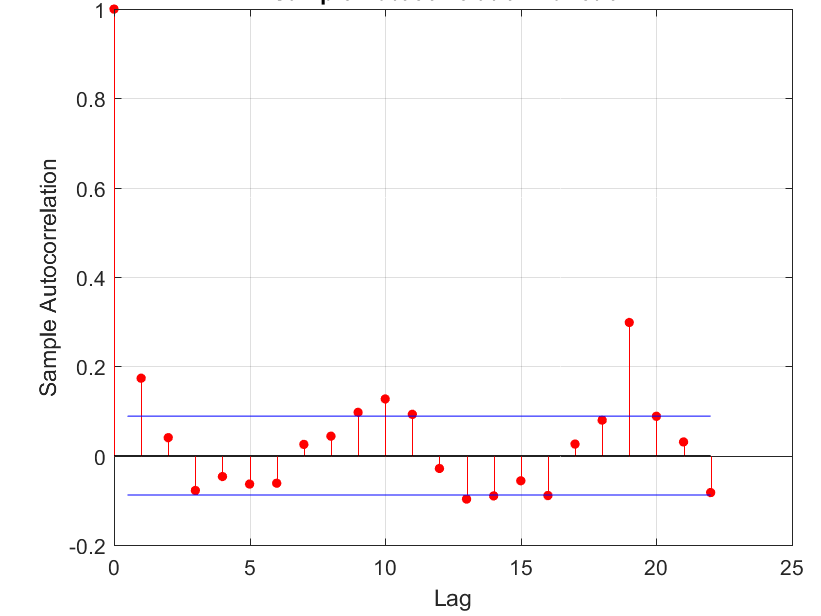}
		\caption{Section 12}
		\label{fig:acf_2}
	\end{subfigure}
	\begin{subfigure}{0.45\textwidth}
		\includegraphics[width=\linewidth]{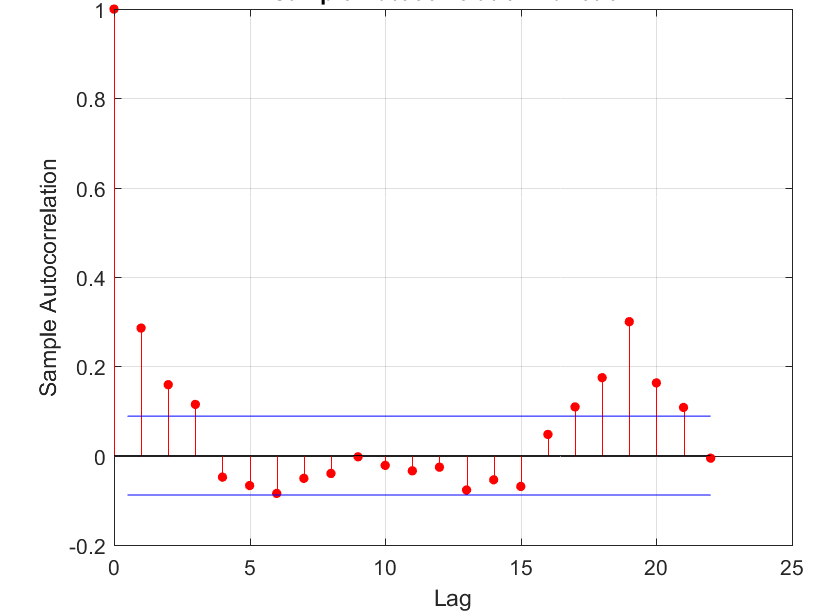}
		\caption{Section 45}
		\label{fig:acf_45}
	\end{subfigure}
	\caption{Sample ACF plots for selected sections of the route}
	\label{fig:acf}
\end{figure}

\begin{figure}[htbp]
	\centering
	\includegraphics[width=0.9\linewidth]{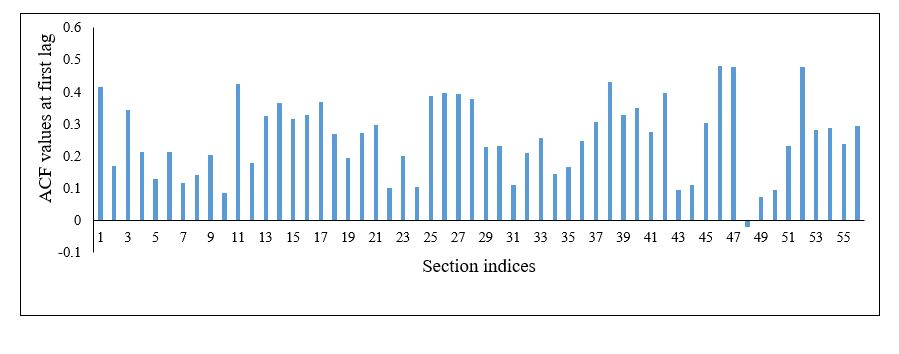}
	\caption{ACF values at first lag for all the sections.}
	\label{fig:lag_1}
\end{figure}

\begin{figure}[h!]
	\centering
	\includegraphics[width=0.9\linewidth]{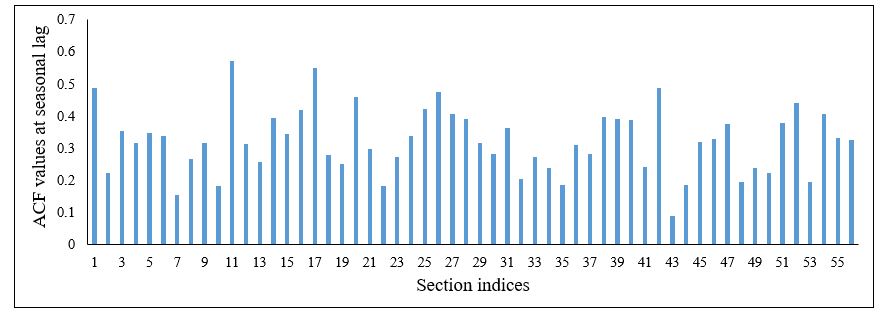}
	\caption{ACF values at first seasonal lag for all the sections.}
	\label{fig:lag_19}
\end{figure}

\noindent A classical (Gaussian distribution based) time series approach for modelling and learning based on historical data and 
subsequent prediction is explained here. Towards the end of the section, we discuss how the lognormality of the data can be incorporated  into the approach 
based on   Gaussian assumptions.
Data from about $27$ days was used for training or model fitting. 
Each day's travel time observations ($19$ of them) at each section were concatenated together (in the order of dates) to form a single long time series.
\FloatBarrier
\vspace{0.25cm}
\noindent {\bf Compensating for standard non-stationarities, if any:} In time series analysis, it is a standard practice to first filter out deterministic
non-stationarities like (polynomial) trends and periodicities (also refered to as seasonality sometimes) from the
data (which is usually non-stationary) before applying standard stationary model fits. On visual inspection, it was observed that there are no linear or higher order trends in the time series of any section. 
Further, on inspection of the Auto Correlation Function (ACF) (sample plots shown in Figure~\ref{fig:acf}), no abnormally slow decay of the ACF was found at consecutive lags. 
Presence of polynomial trends in the data typically manifests in
a slowly decaying ACF which is not the case in our data. Figure ~\ref{fig:lag_1} shows the ACF values at lag $1$ (which are all significantly less than $1$) in all the sections to
ascertain this.  Please observe that the lag $1$ ACF values at all sections are significantly less than $1$, which rules out the presence of polynomial trends.  
Further, since the data has been constructed by stringing together
short time series segments of a fixed length ($19$ in this case), it is natural to check for periodic (seasonal) trends of period one day. The presence of a deterministic seasonal
component can manifest itself in the ACF again with a slow decaying trend at the seasonal lags. Figure ~\ref{fig:lag_19} shows the ACF values at seasonal lag and it can be seen that
the values are significantly less than $1$ in all the sections. Hence, there was no evidence of presence of periodic (seasonal) trends in our data and it was concluded
overall that  deterministic non-stationarities are absent in the data. 

Among the stochastic types of non-stationarity, the well-known integrating type effect was checked using the ADF unit-root
test. We have conducted the ADF test at the optimal lag length to ascertain the presence of unit-root stochastic non-stationarities in the data. First, the maximum lag length for the sample was obtained using Schwert’s thumb rule \cite{sch:89} and ADF test was conducted for all lag lengths. Then, lag length with lowest BIC value is identified as optimal lag length and the corresponding ADF test results were considered for the decision making. In the present study, the ADF test rejected the null hypothesis for all the sections. The integrating type stochastic
non-stationarity can occur at a seasonal level too.  The ACF decay at seasonal lags being significantly less than $1$ is one clear evidence to rule out the presence of seasonal unit
roots. Figure ~\ref{fig:lag_19} which shows the ACF values at the first seasonal lag demonstrates that the decay factor is  significantly less than $1$ in all the sections. This
rules out the presence of seasonal unit roots. 
   
\vspace{0.25cm} 


\begin{figure}[h]
	\centering
	\begin{subfigure}{0.45\textwidth}

		\includegraphics[width=\linewidth]{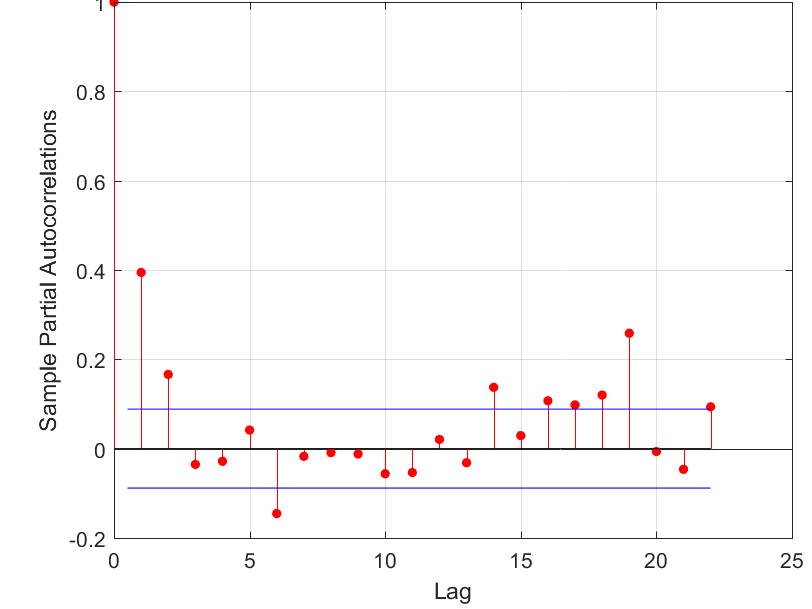}
		\caption{Section 12}
		\label{fig:pacf_14}
	\end{subfigure}
	\begin{subfigure}{0.45\textwidth}

		\includegraphics[width=\linewidth]{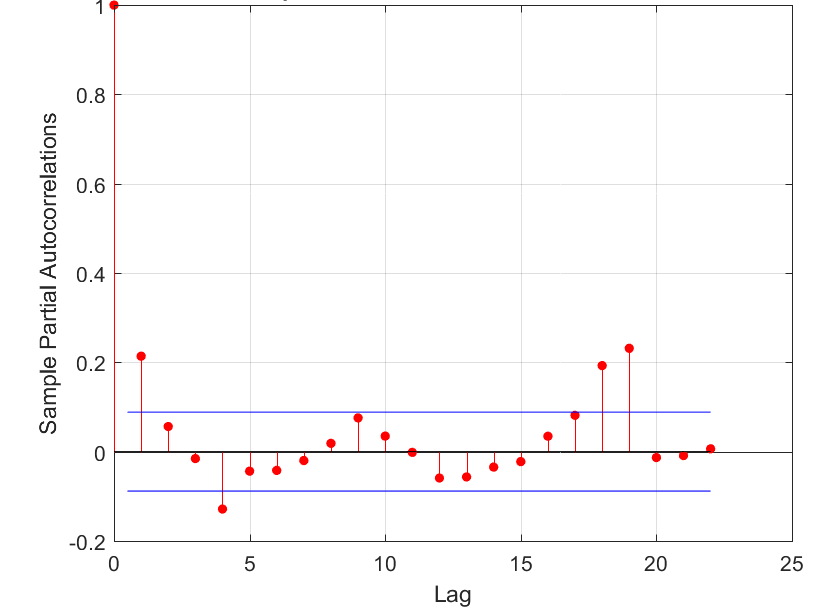}
		\caption{Section 45}
		\label{fig:pacf_40}
	\end{subfigure}
	\caption{Sample PACF plots for selected sections of the route.}
	\label{fig:pacf}
	
\end{figure}

\begin{figure}[htbp]
	\centering
	\includegraphics[width=0.9\linewidth]{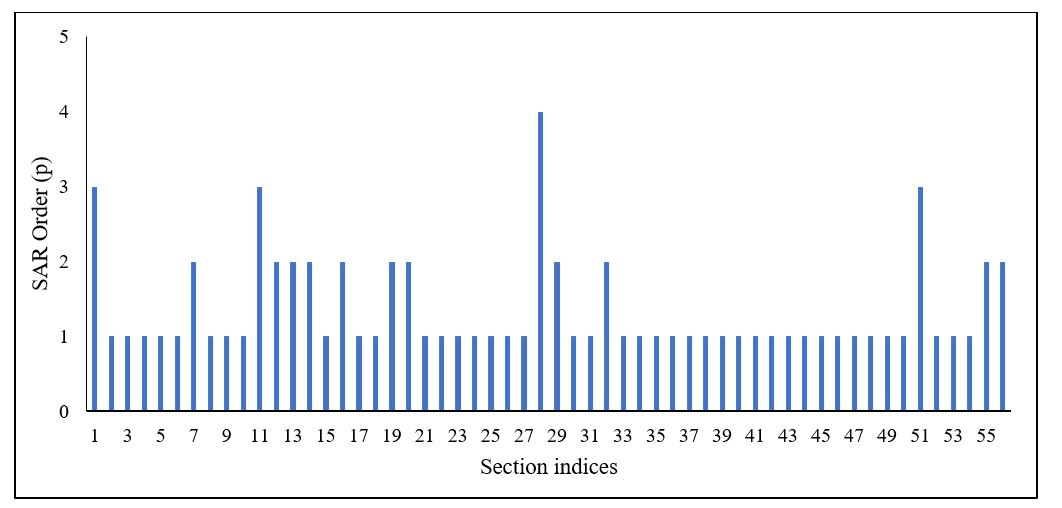}
	\caption{Seasonal AR model order (p) across all the sections.}
	\label{fig:order}
\end{figure}

\noindent {\bf Seasonal AR model fit:} On the conditionally (if the unit-root test detected integrating effects) differenced data, the ACF and Partial Auto Correlation Function (PACF) is computed. 
It was observed that in all cases PACF decayed quickly (shown in Figure~\ref{fig:pacf}).   
This condition is  an ideal candidate
for a pure AR model fit \cite{Box:1990}. Moreover, prediction is a very simple process under AR model fits.
Even though the ACF decays considerably after a couple of initial lags, it again picks up (statistically) significant magnitude  at the intial seasonal lags 
(specifically at $19$ here).  This indicates that even though there are no deterministic seasonal components that repeat in a fixed fashion, there do seem significant 
correlations (stochastic) between travel times exactly a day apart ($19$ ticks here). To incorporate this expected and important feature of the ACF, 
this study propose to use a seasonal AR model 
with an appropriately identified model order using its PACF. It is standard in time series to fix the AR-model order based on the significant PACF co-efficients. This study explores the 
usage of (a) Multiplicative Seasonal AR Model and (b) Additive Seasonal AR Model.  A multiplicative
seasonal AR  model for a process $y(t)$ is of the form 
\begin{equation}
(1-\phi_1L - \phi_2L^2 - \dots -\phi_pL^p)(1-\Phi_1L^s - \Phi_2L^{2s} - \dots -\Phi_kL^{Ps})y(t) =  e(t),
\end{equation}
where, $e(t)$ is a white noise process with unknown variance, $L^p$ is the one-step delay operator applied $p$ times i.e. $L^p{y(t)} =
y(t-p)$.  As the name multiplicative suggests, the AR term in the stationary process is a multiplication of two lag polynomials: (a) first capturing the standard lags of order upto
$p$, (b) second
capturing the influence of the seasonal lags at multiples of the period $s$ and order upto $P$. For instance, for the PACF shown in Fig~\ref{fig:pacf_14}, the choice of order
parameters would be $p=2$, $P=1$, and $s=19$. An additive seasonal AR model on the other hand is a conventional AR model 
$(1-\phi_1L - \phi_2L^2 - \dots -\phi_pL^p)y(t) =  e(t)$
with co-efficients corresponding to the insignificant values in the PACF constrained to be zero. For instance, for the PACF shown in Figure~\ref{fig:pacf_14}, an AR
model of order $19$ is chosen with co-efficients $\phi_3$ to $\phi_{18}$ forced to zero. Both the multiplicative and additive seasonal AR models are learnt using maximum likelihood
estimation \cite{Brockwell:02}. We finally choose the model with the higher AIC between these two for prediction. Figure~\ref{fig:order}  shows the order $p$ of the learnt AR model at every section. Also, this study stuck to one seasonal lag ($P=1$) for building the overall SAR model.  \\
\FloatBarrier
{\bf Incorporate log-normality:} Our first proposed scheme is based on stationary 
log-normal AR
modelling \cite{Stoica}. A log-normal stationary AR process $(Y_1,Y_2,\dots)$ is one whose
transformed process $(X_1,X_2,\dots)$ obtained by applying log on each of its random variables  is a stationary Gaussian AR process. This means $(X_i =
ln(Y_i)), \forall \,i $. As shown in \cite{Stoica}, if one is
predicting  $Y_n$ based on its past values $Y_1,Y_2 \dots Y_{n-1}$, then the conditional distribution of $Y_n$ is also log-normal with parameters
$\mu = \sum_{i=1}^k w_i*X_{n-i}$  and $\sigma^2$. Here $w_i,i=1,\dots k$ are the AR parameters and  $\sigma^2$ is the variance of the input noise, both  in the associated 
Gaussian AR process. 
Accordingly we apply a log transformation on all the travel time observations first, and then fit an AR model as explained above. 
While predicting, in case the data is differenced due to presence of integrating type effects, one needs to do linear prediction on the differenced data and finally undo the 
differencing after linear prediction.    
We finally apply the exponential
transformation to obtain the actual travel time prediction. Applying an exponential transformation is equivalent to choosing the median of the conditional log-normal
distribution as the final point estimate which is statistically optimal under the  mean absolute error loss function \cite{Kay:1993}. Log-normal modelling (unlike Gaussian modelling) gives us the flexibility of trying out the mean and 
mode (which are optimal under the 
mean square error and  $\epsilon$-insensitive loss functions \cite{Kay:1993} as well apart from the median (all three of which are distinct)  for the final point estimates. 
Both these point estimates additionally would need the conditional variance of the prediction in the Gaussian domain.   
 
\subsection{Linear  Non-Stationary Approach}

The classical time series approach discussed before assumes the data to be a sum of possible trend and periodic (deterministic) non-stationarities plus a stochastic stationary component 
(with possible integrating type
stochastic non-stationarities). The next approach that is discussed here, on the other hand  models the data in a non-stationary fashion directly. The conventional time-series approach
(discussed before) assumes one long
realization of a random process. It assumes ergodicity of the process for estimating ensemble averages using time averages \cite{Brockwell:02}. However, the non-stationary model we
now consider involves a fixed finite number 
of random variables. The number of random variables in this random vector will be equal to the number of slots we bin the 24 hr axis into, at each section. 
\comment{
	This approach tries to directly model (at each section) 
	the travel time evolution of
	buses across the time-bins in a day without assuming any stationarity on the data. In this sense, the model involves only a finite number of random variables equal to the number
	of time-bins in a day. }
To learn the statistics of these finite set of random variables, one needs independent and identically distributed (i.i.d.) realizations of this finite-dimensional random vector. The travel time vector (at a fixed section)  of a particular day which consists of observations from different time-bins of the day can be regarded as one realization of 
this random vector. The collection of all such travel time vectors across all training days constitute i.i.d. realizations of this random vector.

Recall from Figure~\ref{fig:lag_1} of the previous (SAR) subsection that we did not find any 
trend in the travel time
sequential (time-series) vector  which was obtained by concatenating all the daywise travel time training vectors in order. This justifies our assumption   to regard the daywise travel time
vectors at a particular section  to be identically distributed. However, this joint distribution can change from section to section. Presence of a trend in the day-wise 
concatenated time series would nullify our assumption that these day-wise travel time vectors are identically distributed.

\begin{figure}[h!]
	\centering
	\begin{subfigure}{0.49\textwidth}
		\includegraphics[width=\linewidth]{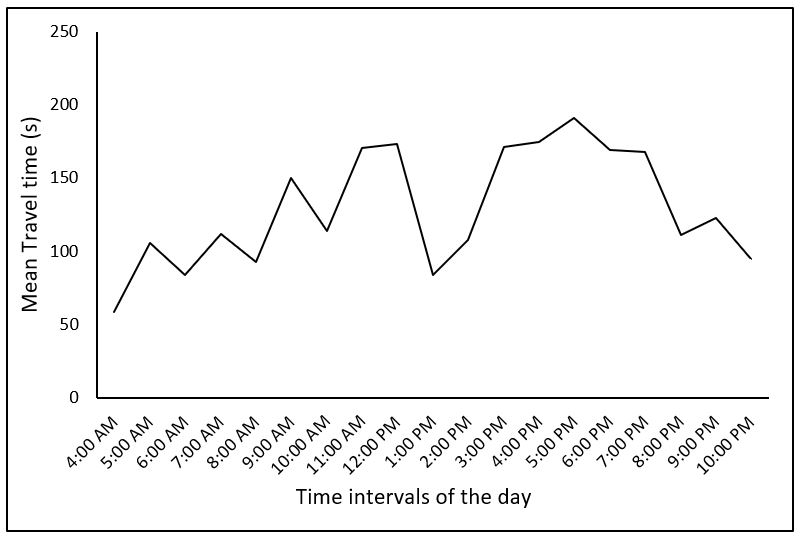}
		\caption{Section 27}
		\label{fig:mean_1}
	\end{subfigure}
	\begin{subfigure}{0.49\textwidth}
		\includegraphics[width=\linewidth]{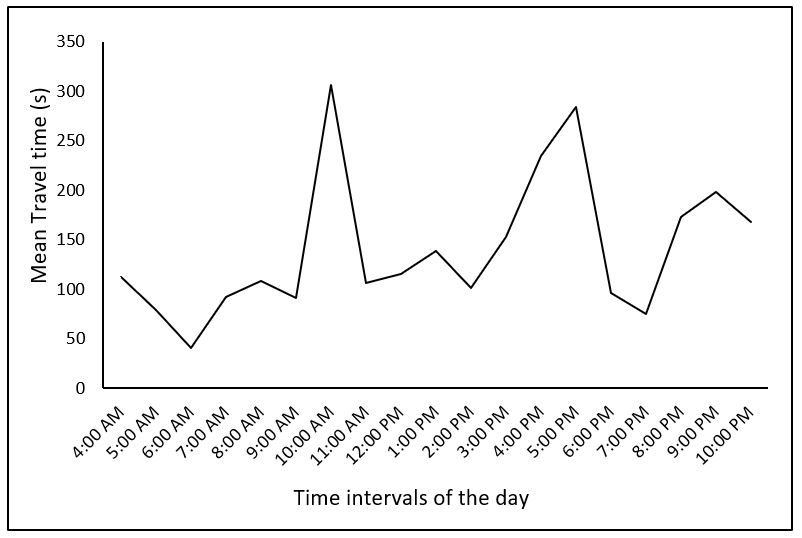}
		\caption{Section 47}
		\label{fig:mean_2}
	\end{subfigure}
	\caption{Variation of mean travel time across time intervals of the day}
	\label{fig:meanplot}
	
\end{figure}

\begin{figure}[htbp]
	\centering
	\includegraphics[width=\linewidth]{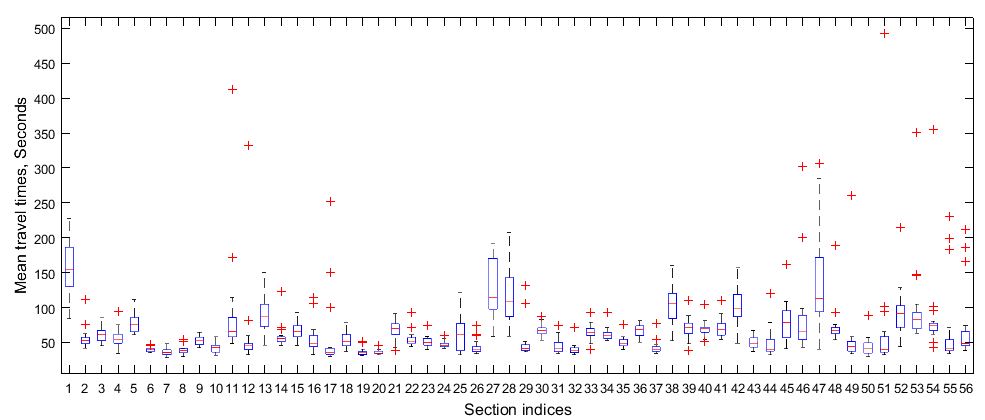}
	\caption{Box plot of mean travel time across sections.}
	\label{fig:boxplot}
\end{figure}

Hence, it can be assumed that these 27 travel time vectors as i.i.d and estimate the joint distribution of these 19-dimensional random vector at each section
separately. Before we proceed to fit a predictive model, we first look at the empirical mean of each of these 19 random variables at every section. We provide the mean variation
plot at two randomly chosen sections across the 19 time bins in Figure \ref{fig:meanplot}. We
further provide a box plot of the mean travel times in a day across all the 56 sections in Figure \ref{fig:boxplot}. This overall justifies the evidence for (mean) non-stationarity at each section amongst these $19$ random variables. 

\FloatBarrier

Motivated from this observation, the proposed approach tries to fit a general  Gaussian/ log-normal model to aid real-time predictions based on the travel-time of the past 
time-bins of the same day. We propose to use a  non-stationary auto-regressive modelling approach, 
which makes prediction a straightforward and computationally inexpensive task.
Firstly,  the non-stationary AR-model is explained under the Gaussian assumption. Extending it further under a log-normal joint distribution is similar to the stationary log-normal AR case. 

\begin{figure}[htbp]
	
	\includegraphics[width=0.9\linewidth]{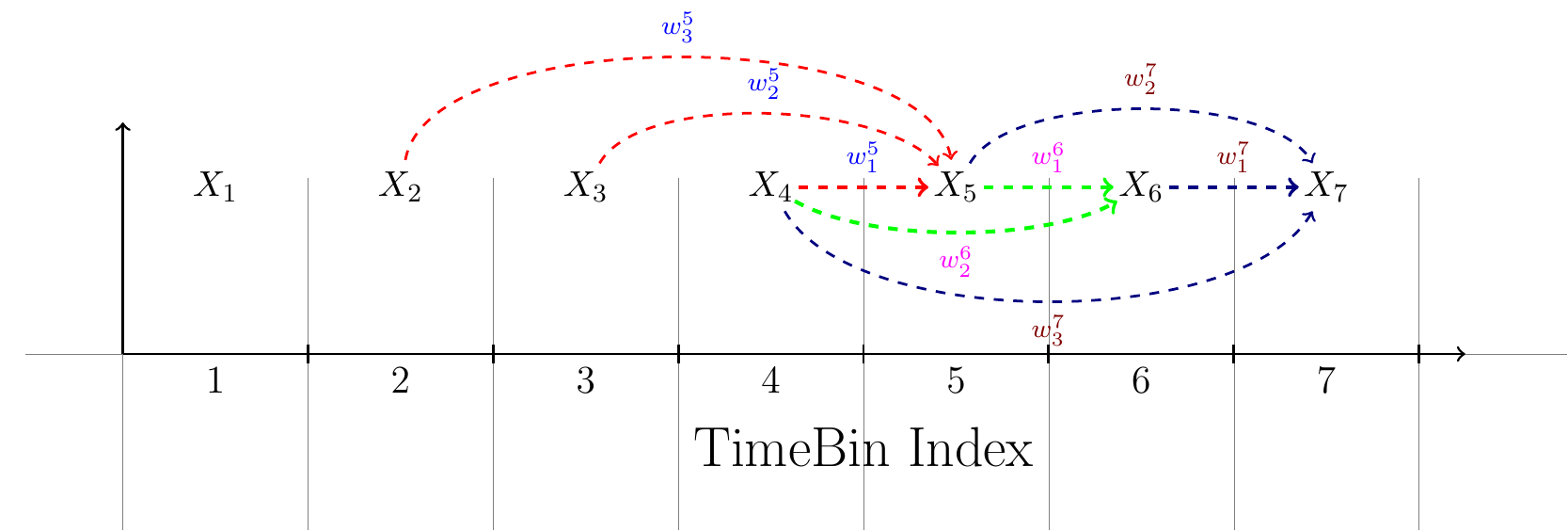}
	\caption{Illustration of non-stationary model: Influences on to a particular time bin indicated via arcs of the same color and associated weights $w_i^n$ (except $w_0^n$) at 
		time bin $5$, $6$ and 
		$7$. $k(5)=3$, $k(6)=2$, $k(7)=3$.}
	\label{fig:NS}
\end{figure}

As the name auto-regressive suggests, the proposed model assumes that any travel time observation  can be obtained by linearly regressing a finite set of preceding past 
values. In particular, any travel time observation $X_n$ (at time-bin $n$) can be 
written as a linear combination (with a bias) of a set of its $k(n)$ immediately preceding 
values plus some independent Gaussian noise with an unknown time-varying ($n$-dependent)  variance. \comment{
	We propose another interesting approach where we tackle the non-stationarity of the data in a different way. In this approach, we assume the data follows a non-stationary
	Gaussian AR process. By this, we mean that the any random variable at time $t$ can be obtained by linearly regressing with a finite set of immediately preceding past values. 
} This form of  non-stationarity brings in two levels of generality as opposed to a stationary Gaussian AR process: (a) The regressing weights now would be a function of $n$, (b) The number of immediately preceding past (travel-time) values that influence at  time-bin $n$ can also vary with $n$ i.e. the weight vector length $k(n)$ is now a function of $n$. 
We denote the weight vector at time bin $n$ as $\mathbf{w}^n = [w_0^n,w_1^n,w_2^n,\dots
w_{k(n)}^n]$. Figure~\ref{fig:NS} illustrates these  non-stationarity influences.

At each section, the sequence of travel time observations of each historical day is considered a
realization of the above non-stationary process. For each section, the entire  collection of all historical (27) days is treated as (27) i.i.d realizations from 
this process. 
The availability   of i.i.d. data from many days in the historical past makes estimation of the time-varying regression weight vectors $\mathbf{w}^n = [w_0^n,w_1^n,w_2^n,\dots
w_{k(n)}^n]$ feasible for each time-bin $n$. The estimation process can be a little tricky given that the length of
the weight vector is not known apriori.  The problem boils down to solving the following statistical conditional independence question: 
Given the recent past $k$ values upto time $n-k$, are the current and the past beyond time $(n-k)$ conditionally independent? If this is the case, the main advantage is as follows. To 
predict the current travel time $X_n$, knowledge of only its $k$ past values are necessary and one can forget about the past beyond (or before) $X_{n-k}$. 
We are in particular interested in the least such $k$ for which
this holds. Checking this conditional independence for general distributions can be complicated or even infeasible. However for Gaussian distributions, the notion of partial correlation
(PC), comes to our aid here, which is defined as follows.

Partial correlation as the name suggests captures the  correlation between those parts of $A$ and $B$ that are not explainable by  $\mathbf{C}$.  Interestingly, for multi-variate
Gaussian distributions, it turns out that 
the conditional independence between $A$ and $B$ given $\mathbf{C}$ holds if and only if the associated PC between $A$ and $B$ given $C$ is $0$  \cite{baba04},  as
stated next. This crucial relation is what facilitates a method to ascertain the necessary conditional independence to build our non-stationary auto-regressive model. 

\begin{definition}
	\label{def:PC}
	\cite{anderson:03,shumway:05} Given random variables $A$ and $B$ and a random vector $\mathbf{C}=(C_1,C_2,\dots C_n)$, the partial correlation between $A$ and $B$ given $C$ is 
	defined as the correlation coefficient between residual errors $e_A = (A - f_1(\mathbf{C}))$ and $e_B = (B - f_2(\mathbf{C}))$, where  $f_1(\mathbf{C})$ is the optimal linear predictor (in
	the mean square sense) of $A$ given $\mathbf{C}$ and $f_2(\mathbf{C})$ is the optimal linear predictor (in
	the mean square sense) of $B$ given $\mathbf{C}$.
\end{definition}

\begin{property}
	Consider random variables $X$ and $Y$ and a random vector $\B{Z}$ such that $(X,Y,\B{Z})$ are jointly Gaussian. $X$ and $Y$ are conditionally independent given $\B{Z}$ if
	and only if the partial correlation between $X$ and $Y$ is zero.
\end{property}

\noindent \textit{\textbf{Proof.}} For Gaussian joint distributions, all conditional distributions are also Gaussian with mean being a linear function of the conditioning random variables while the variance
	is constant and independent of the values assumed by the conditioning random variables \cite{rohatgi15}. This means the joint distribution of $X$ and $Y$ (conditioned on
	$\B{Z}$) would be Gaussian. It is an elementary fact that independence of two jointly Gaussians is 
	equivalent to their
	covariance being zero.   Hence, checking for conditional independence of $X$ and $Y$ is equivalent to checking for their conditional
	covariance being zero. Checking for conditional variance being zero for every value of the conditioning random variables is in general computationally infeasible. However,
	in the joint Gaussian case, one can show that the conditional variances and conditional correlation co-efficients are independent of the specific values the conditioning
	random variables can assume \cite{kendall77}. This means that the conditional covariance which is the product of conditional variance and conditional correlation
	co-efficient will be independent of the conditioning random variables. 
	
	The linear regressor of $X$ w.r.t. $\B{Z}$ i.e. the best linear approximation of $X$ in terms of $\B{Z}$ in the mean-square error (MSE) sense is $E(X/\B{Z})$ when 
	$X$ and $Z$ are jointly Gaussian. For similar reasons, the linear regressor of  $Y$ w.r.t $\B{Z}$  is $E(Y/\B{Z})$.
	Hence in the multi-dimensional Gaussian setting, the partial correlation can be written as 
	\begin{eqnarray}
	&&E\left[(X-E(X/\B{Z}))(Y-E(Y/\B{Z}))\right] \\ \nonumber
	&=& \int E\left[(X-E(X/\B{Z}))(Y-E(Y/\B{Z}))/\B{Z} = \B{z}\right]f_{\B{Z}}(\B{z}) d\B{z}  \\ \nonumber
	&=& \int E\left[(X-E(X/\B{Z}= \B{z}))(Y-E(Y/\B{Z}= \B{z}))/\B{Z} = \B{z}\right]f_{\B{Z}}(\B{z}) d\B{z},
	\end{eqnarray}
	
	\noindent the first term in the above integral is the conditional covariance between $X$ and $Y$ which is independent of $\B{z}$ as explained above. Hence this constant term can
	come out of the integral and PC is now equal to this constant conditional covariance.  Hence the property follows.

\comment{
	Owing to this simplification in the Gaussian setting, one can show by a simple conditioning argument that the constant 
	conditional covariance is equal to the PC \cite{baba04}. Hence checking for PC being zero is sufficient. 
}
\begin{algorithm}[t]	
	\caption{Learn/Estimate AR Co-efficients and Noise Variance at Each Time Bin $n$ {\em for a Fixed Section.}}
	\label{algo:prediction}
	\KwIn{Historical Data $\scr{D} = [\B{x_1}\,\B{x_2}\,\dots\, \B{x_T}]_{(d\times S)}$, $S$ - total no. of time bins in the day. $d$ - number of days for training, 
		$\B{x_i}_{(d\times 1)}$ - travel time observation vector at the $i^{th}$ time bin of the day. 
	}
	\KwOut {$\mathbf{w}^n = [w^n_0,w_1^n,w_2^n,\dots w_{k(n)}^n]$, $\sigma_n^2$.}
	Initialize $\Fl=1$\; 
	\For{\IW  $\lA (n-1)$ \KwTo $2$ }{
		Linearly regress $X_{n}$ w.r.t $(X_{n-1},X_{n-2},\dots X_{\IW})$ to obtain $\B{w^f}$ with  INPUT = $(\B{x_{n-1}},\B{x_{n-2}},\dots \B{x_{\IW}})$, OUTPUT =
		$(\B{x_{n}})$. \\
		Linearly regress $X_{\IW - 1}$ w.r.t $(X_{n-1},X_{n-2},\dots X_{\IW})$ to obtain $\B{w^b}$ with  INPUT = $(\B{x_{n-1}},\B{x_{n-2}},\dots \B{x_{\IW}})$, OUTPUT =
		$(\B{x_{\IW-1}})$. \\
		From 	$\B{w^f}$ and $\B{w^b}$, compute the respective sample residuals $\B{r_f}$ and $\B{r_b}$, as below: \\
		$\forall i,\, \B{r_f}(i) \lA \B{x_n}(i) - \sum_{j=1}^{n - \IW} \B{w^f}(j)*\B{x_{n-j}}(i) + \B{w^f}(0) $, $i$ - day index. \\
		$\forall i,\, \B{r_b}(i) \lA \B{x_{\IW}}(i) - \sum_{j=1}^{n - \IW} \B{w^b}(j)*\B{x_{n-j}}(i) + \B{w^b}(0) $, $i$ - day index. \\
		Compute \PC, sample correlation coefficient between $\B{r_f}$ and $\B{r_b}$. \\
		/* The above computes partial Correlation between $X_n$ \& $X_{\IW - 1}$ given the intermediate random variables $(X_{n-1},X_{n-2},\dots X_{\IW})$. */	\\
		Compute test statistic $t^* \lA \frac{\PC\sqrt{d-2}}{1 - \PC^2}$  to assess if $\PC==0$ based on a standard t-test (\cite{kendall77}). 		\\ 
		\If{p-value of $t^* > 0.05$}{
			$\Fl = 0$ \;
			break\;
		}
	}        
	\If{$\Fl == 1$}{
		Linearly regress $X_{n}$ w.r.t $(X_{n-1},X_{n-2},\dots X_{1})$ to obtain $\B{w^f}$ with INPUT = $(\B{x_{n-1}},\B{x_{n-2}},\dots \B{x_{1}})$, OUTPUT =
		$(\B{x_{n}})$. \\
		Compute sample residual $\B{r_f}$ from $\B{w^f}$.
	}
	$\sigma_n^2 \lA $ Sample variance of $\B{r_f}$ \; 
	\Return{$(\B{w^f},\sigma_n^2)$} 
\end{algorithm}

Algorithm~\ref{algo:prediction} gives a detailed explanation  of computing the order and values of auto-regression and the additive noise variance 
for a specific time bin $n$.  The algorithm basically computes the sample partial correlation (PC) between $X_n$ and $X_{win - 1}$ given all the intermediate
random variables namely $(X_{n-1},X_{n-2},\dots X_{\IW})$  for  $win$ varied progressively from $n-1$ to $2$ (line $2$). For a particular $win$, it performs a forward regression to 
compute $\B{w^f}$, the optimal linear
predictor of $X_n$ in terms of $(X_{n-1},X_{n-2},\dots X_{\IW})$ (line $3$). Similarly, it performs a backward regression to compute $\B{w^b}$, the optimal linear
predictor of $X_n$ in terms of $(X_{n-1},X_{n-2},\dots X_{\IW})$ (line $4$). The associated residual errors of both the forward and backward linear regressions are computed next
(lines $5$ to $7$). As per definition~\ref{def:PC}, sample PC is computed by calculating the correlation coefficient of the $\B{r_f}$ and $\B{r_b}$ (line 10). To statistically ascertain 
if the computed sample PC is zero or not, we use hypothesis testing. In particular, our null hypothesis used is PC equal to zero.  We employ  a standard t-test \cite{kendall77} for 
assessing zero correlation coefficient with test statistic $\frac{\PC\sqrt{d-2}}{1 - \PC^2}$ (line $10$) which is known to follow a $t$-distribution with $d-2$ degrees of freedom. 
If the $p$-value of the test statistic is less than $0.05$ (significance level), we reject the null-hypothesis. Note that the test is 2-sided with the test statistic possibly taking
both positive and negative values. We stop at the earliest instance of $win$ where $p$-value is greater than $0.05$ (line $11$ to $13$). This means we retain the null hypothesis and the 
partial  correlation is assessed to be zero. This would give us the order of autoregression, $k(n)=(n-win)$ at time bin $n$. 
This is also the minimum number  of previous travel times  conditioned on which the current $X_n$
and the past ($X_{win - 1}$ and beyond)  are independent.  
If the PC is
not significantly close to zero for any $win$,  then $X_n$ is dependent on the entire past (lines $16$ and $17$).

{\bf Incorporating lognormality:} A lognormal non-stationary AR process can be defined as one whose log-transformed process is a Gaussian non-stationary AR process. Learning under this model, can
be readily achieved by first taking the log of all the observations and then fitting a gaussian non-stationary AR model as explained above. 

The optimal prediction for the Gaussian non-stationary AR model at time tick $n$ given the current past $(X_1,X_2 \dots X_{n-1})$ is carried out as 
$\hat{X}(n) = w^n_0 + \sum_{i=1}^{k(n)} w^n_i*X_{n-i}$. In the lognormal case, the original current observations $(Y_1,Y_2 \dots Y_{n-1})$ are log transformed first to
obtain $(X_1,X_2 \dots X_{n-1})$. The optimal linear combination in the Gaussian domain is computed like earlier $\hat{X}(n) = w^n_0 + \sum_{i=1}^{k(n)} w^n_i*X_{n-i}$
and the final optimal point estimate in the lognormal domain is $\hat{Y}(n)= exp^{\hat{X}(n)}$. This point estimate turns out to be the median of the conditional
distribution $P(Y_n/Y_{n-1},Y_{n-2}\dots Y_1)$, which is also lognormal. One can also use the mean and mode of this conditional lognormal distribution for a point estimate 
in which case we would additionally need
information of $\sigma_n^2$ (in addition to $\hat{X}(n)$).
However for the
purposes of the current paper we stick to the median which translates to the use of a simple exponentiation of the prediction in the Gaussian domain. 

\FloatBarrier

\subsection{Real-time prediction across multiple sections ahead}
In the previous two subsections, linear statistical models (with lognormality incorporated while not necessarily stationary) which capture temporal correlations in the bus 
travel time data were proposed. 
The temporal prediction models  built  at each section need to be ultimately used for real-time predictions of travel times for a bus whose trip is currently active.  
The constructed models can predict
temporally ahead into the future at every section. However, what we need is predicting travel times of a currently plying bus over a couple of sections ahead of its current position. It is 
not immediately clear how the sequential prediction models developed in the previous sections could be used for travel-time prediction over a subroute of arbitrary sections ahead 
from the current position of an active bus.
\comment{
	\begin{algorithm}[t]
		\caption{Travel-Time Prediction across $k$-sections ahead}
		\label{algo:MultiStepPrediction1}
		
		\KwIn{Current Real Time Data of all previous buses, Current position of the current bus at the end of some section $i$, Current Time $T_{curr}$.  
		}
		\KwOut {Predicted Travel Times to travel $k$ sections ahead from section $i+1$ starting at time $T_{curr}$.}
		Compute the time bin, \CTB within which $T_{curr}$ falls \;
		Initialize $T_{acc}  = T_{curr} -$ StartTime(CurrBin) \;
		Initialize \TSPI = 1 \;
		\For{$j=1,2,\dots k$}{
			/* Predict Travel time at section $i+j$ using the current real time data at section $i+j$ upto time bin \CTB\,- 1. */ \\
			/* In other words, predict travel time for time bin  \CTB +  (\TSPI\,- 1) at section $i+j$.*/  \\
			Do \TSPI-{\bf step} temporal linear prediction using either the seasonal AR model OR the non-stationary AR model in the Gaussian domain (using the log transformed raw data) 
			to obtain $T_{pred(i+j)}$. \\
			$TravelTime = TravelTime +  exp(T_{pred(i+j)})$. \\
			$T_{acc} = T_{acc} + exp(T_{pred(i+j)})$. \\
			\If{$T_{acc}\geq \Delta$}{
				\TSPI =  \TSPI + 1\;
				$T_{acc} = T_{acc} - \Delta$\; 
			}
		} 
		\Return{$TravelTime$} 
	\end{algorithm}
}

\begin{figure}[htbp]
	
	\includegraphics[width=1\linewidth,height=0.7\linewidth]{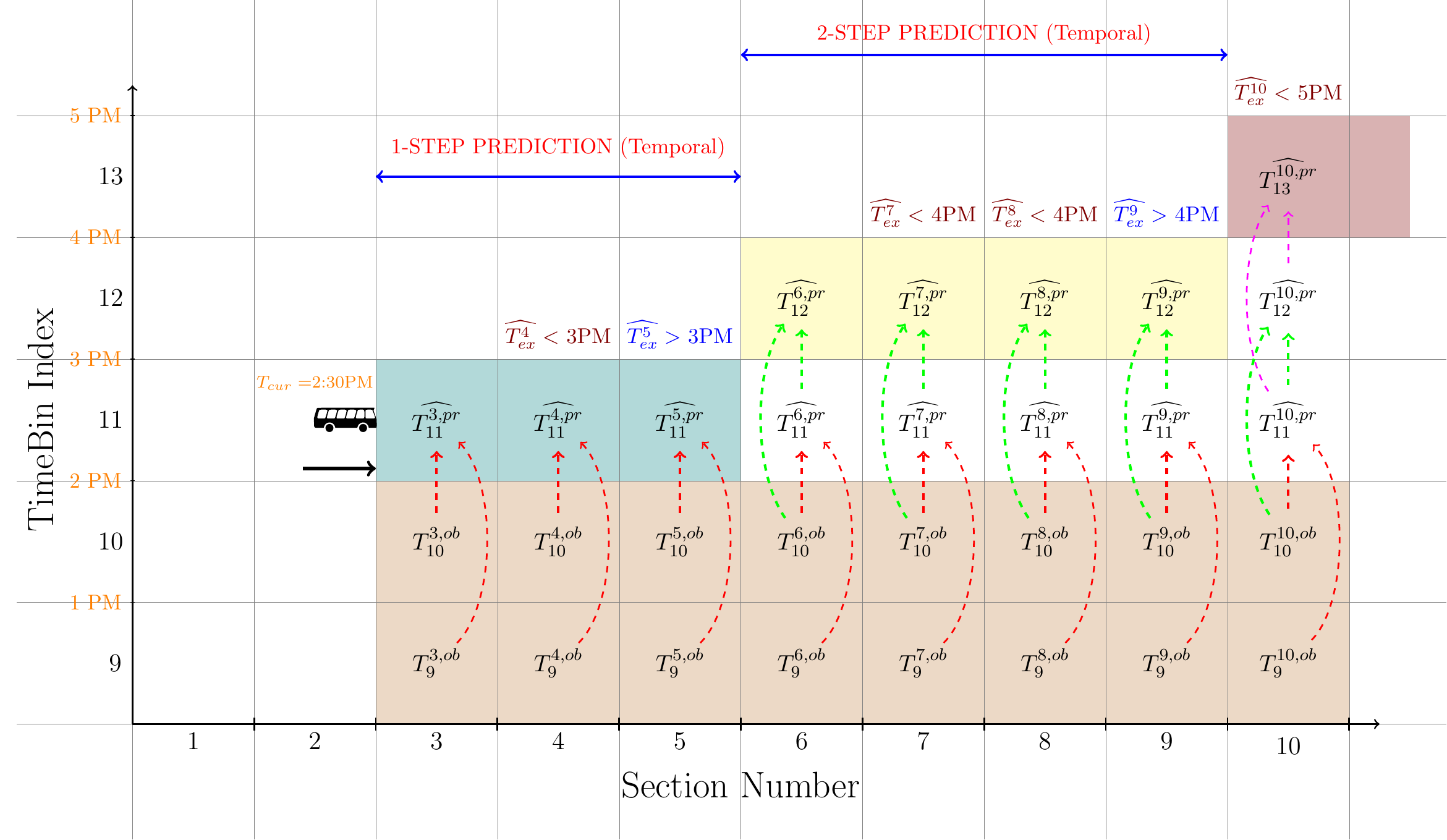}
	\caption{Illustration of the multi-section ahead travel time prediction.}
	\label{fig:MultiSection}
\end{figure}

We illustrate the method via an example as shown in the space-time diagram of  Figure~\ref{fig:MultiSection}. We assume that the time bins are all of the same duration of one hour
even though our proposed methodologies can be used for general non-uniform time bin durations. 
The bus is currently on the verge of leaving current section $i$ ($i=2$ here). The current time is $2:30$ PM with
the current time bin $j=11$ in this example. We assme that the   
current real-time information is available on these various subsequent sections upto the previous time bin ($j-1=10$). The current (or observed) real-time travel time information 
at a subsequent section
$a$ and time bin $b$ is denoted as
$T^{a,ob}_b$, while the predicted travel time at a section  $a$ and time bin $b$ is denoted as  $\widehat{T^{a,pr}_b}$.
In the example, it is further assumed that the learnt models are simple autoregressive models of order $2$ for ease of illustration. This means we need real-time data from only 
the previous two time bins as the shaded region of the first two rows of the space-time diagram indicates. 
Prediction of travel time on  one section ahead of the bus's current position (section $3$ in the example) is pretty straight forward. 

Given that the bus is currently just leaving section  $i$ at time bin $j$ ($j=11$ in the example), to predict the bus's travel time on section $i+1$, one needs to use 
the current
real-time travel
time information on section $i+1$ (from all previous buses) upto previous time bin $j-1$. One needs to perform a one-step temporal prediction at section $i+1$ to predict the bus's
travel time one section ahead as shown in Figure~\ref{fig:MultiSection}. 

To extend this to an arbitrary number of sections ahead in a principled way, one would need to potentially perform a {\em multi-step  temporal} prediction 
(using either of the sequential models previously discussed) at the subsequent
sections. For a subsequent section $\ell$, the larger the difference between $\ell$ and current section $i$, the number of temporal steps ahead we need to predict at section $\ell$ will
proportionately go up. The fundamental idea of the algorithm is as follows. It sequentially goes through the sections ahead starting from section $i+1$. We predict the travel times based on the temporal model learnt at
that section by performing a one-step prediction to start with. For every travel time predicted at a subsequent section $k$, we update the expected exit time of the bus from section
$k$ by adding this prediction
time. The exit time thus calculated for section $k$ is also the expected entry time into section $k+1$. {\em The first instance when the expected exit time enters the next time bin
	(namely $j+1$), the 
	temporal prediction from the next section onwards has to be $2$-step ahead.} Note that this happens at section $5$ in the illustration of Fig.~\ref{fig:MultiSection} upto which a
single-step prediction is performed to calculate the estimates as clearly indicated via the quantities $\widehat{T^{3,pr}_{11}}$, $\widehat{T^{4,pr}_{11}}$ and $\widehat{T^{5,pr}_{11}}$. 
A $2$-step temporal prediction is adopted from the next section because the
current real-time data we have from the previous buses is only upto time bin $j-1$, while the bus would reach this subsequent section only in time bin $j+1$. Accordingly from now on, 
for a consecutive block of sections one performs a two-step prediction till the first instance
the expected exit time moves into time bin $j+2$ and the process continues. In the example, the next such block of sections where $2$-step prediction is carried out ranges from $6$
to $9$.

\begin{algorithm}[t]
	\caption{Travel-Time Prediction Across $k$-sections Ahead}
	\label{algo:MultiStepPrediction}
	\KwIn{Current Real Time Data of all previous buses, Current position of the current bus at the end of some section $i$, Current Time $T_{cur}$.  
	}
	\KwOut {Predicted Travel Times to travel $k$ sections ahead from section $i+1$ starting at time $T_{cur}$.}
	Compute the current time bin, $j$ within which $T_{cur}$ falls \;
	Initialize $\widehat{T_{ex}}  = T_{cur} $  \;
	Initialize \TSPI = 1 \;
	Initialize \ETB = $j$ \;
	\For{$k=1,2,\dots K$}{
		Do \TSPI-{\bf step} temporal linear prediction at section $i+k$ using either the seasonal AR model OR the non-stationary AR model in the Gaussian domain (using the log transformed raw data) 
		to obtain $T^{(i+k),pr}$. \\
		/* Predict Travel time at section $i+k$ using the current real time data at section $i+k$ upto time bin $j$\,- 1. */ \\
		/* In other words, predict travel time for time bin  $j$ - 1 + (\TSPI) at section $i+k$.*/  \\
		$\widehat{T_{ex}} = \widehat{T_{ex}} +  exp(T^{(i+k),pr})$. \\

		\If{$\widehat{T_{ex}}$ falls outside \ETB}{
			\TSPI =  \TSPI + 1\;
			\ETB =  \ETB+ 1\;
		}
	}        
	\Return{$TravelTime$} 
    
\end{algorithm}

Algorithm~\ref{algo:MultiStepPrediction} presents a pseudocode for achieving this. 
Current time, 
$T_{cur}$ happens to fall in the current timebin, $j$,  numbered $11$ with duration $2$ pm to $3$ pm (line $1$). $\widehat{T_{ex}}$ captures the dynamically predicted exit time from a section as the 
algorithm iterates
sequentially through subsequent sections. \TSPI captures the number of time-steps ahead that the temporal models should predict as the algorithm proceeds. \TSPI starts with $1$
(line $3$). \ETB captures the time bin in which the bus is expected to be as the prediction proceeds forward sequentially in space. It is as expected initialized to $j$, the current
time bin (line $4$). 
\TSPI gets incremented suitably whenever the exit time $\widehat{T_{ex}}$ falls beyond the \ETB (line $11$). \ETB also gets incremented under the same condition as shown in line
$12$. At the $k^{th}$ subsequent section (from section $i$), we make a $\TSPI$-step ahead prediction of the temporal model pre-learnt at section $i+k$ (line $6$).
Fig.~\ref{fig:MultiSection} clearly
illustrates how the number of temporal steps ahead prediction needed at a section $i+k$ is proportional to $k$.  It starts with a consecutive block of sections on which the temporal
models perform a $1$-step temporal prediction followed by the next consecutive block of sections, in each of which a $2$-step temporal prediction is carried out and so on.
\comment{
	The travel time prediction at the first couple of sections beyond its current position at section $i=2$ would be performed  for the current time bin at $2$ to $3$ pm as the query time of 
	prediction is still 2:30 pm.  
	Further, there will come a point of transition when the
	total (predicted) travel time across the first couple of sections exceeds $30$ minutes (End Time of Current Time Bin (3 pm) - Time of Prediction
	Query (2:30 pm)) and the bus 
	is now effectively in the time zone of time bin
	$3$ to $4$ pm. For this
	next block of sections, the travel time prediction needs to be carried out for the time bin $3$ to $4$ pm. Our current real time data that is available till now is
	effectively till $2$ pm (end of the last time bin before the time of prediction query) from the previous buses. This means to predict travel time at these sections for the time bin $3$ to $4$ pm, one needs to perform a two-step temporal
	prediction at each of these sections. In general, the variable \TSPI (initialized in line $3$) precisely captures how many time steps ahead should prediction be carried out using the 
	learnt temporal
	correlation models at each section in such a block of sections.  Line $7$ captures the variable time step ahead prediction that is carried out
	depending on the dynamic value of  \TSPI.   A block of consecutive sections (where the associated predictors operate at the same multi-step
	ahead temporal prediction)  gets terminated when the sum of predicted travel times across these
	consective sections exceeds the time bin size ($\Delta$) for the first time (line $10$). The variable $T_{acc}$ keeps track of these accumulating sum of travel times in each such
	growing block. The variable $Travel Time$ keeps track of the sum of  predicted travel times  across all currently processed sections and finally returns the total
	predicted travel time across $k$-sections ahead.
}

\section{RESULTS}

\noindent  \noindent  The proposed methodologies were implemented in MATLAB for both single step prediction and multi step prediction and the performance was evaluated in terms of Mean Absolute Percentage Error (MAPE) and Mean Absolute Error (MAE). Testing was carried out for a period of seven days and results are discussed below. 
\vspace{-0.5cm}
\subsection{Performance Evaluation }
\noindent The present study evaluated the predicted travel time values with the actual measured travel time values. The evaluation was done at three stages to highlight the observations and contributions of the present study. First part of the study evaluates the log-normal assumptions of the data by comparing the results of Seasonal AR method and  Non-stationary method under both Gaussian and log-normal assumptions at an aggregate level. Next, the proposed methods were compared with each other and the inferences were made at section level and trip level to know the superiority amongst them. Finally, the performance of the multi-step ahead prediction framework was evaluated and compared between the Seasonal AR and the non-stationary method.

\FloatBarrier

\subsubsection{Single step prediction}
\begin{figure}[h]
	\centering
	\begin{subfigure}{0.9\textwidth}
		\centering
		\includegraphics[width=0.65\linewidth]{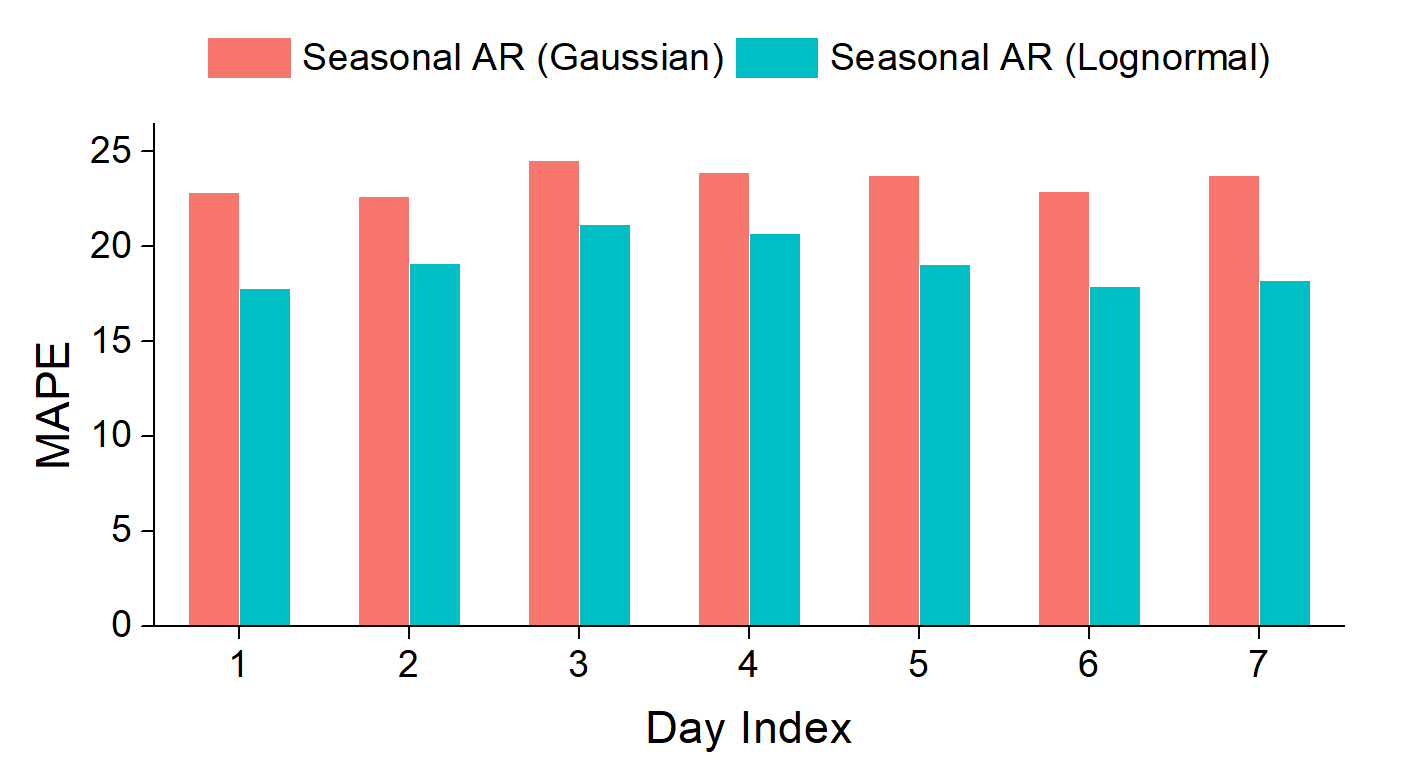}
		\caption{Seasonal AR method}
		\label{fig:log_SAR}
	\end{subfigure}
	\begin{subfigure}{0.9\textwidth}
		\centering
		\includegraphics[width=0.65\linewidth]{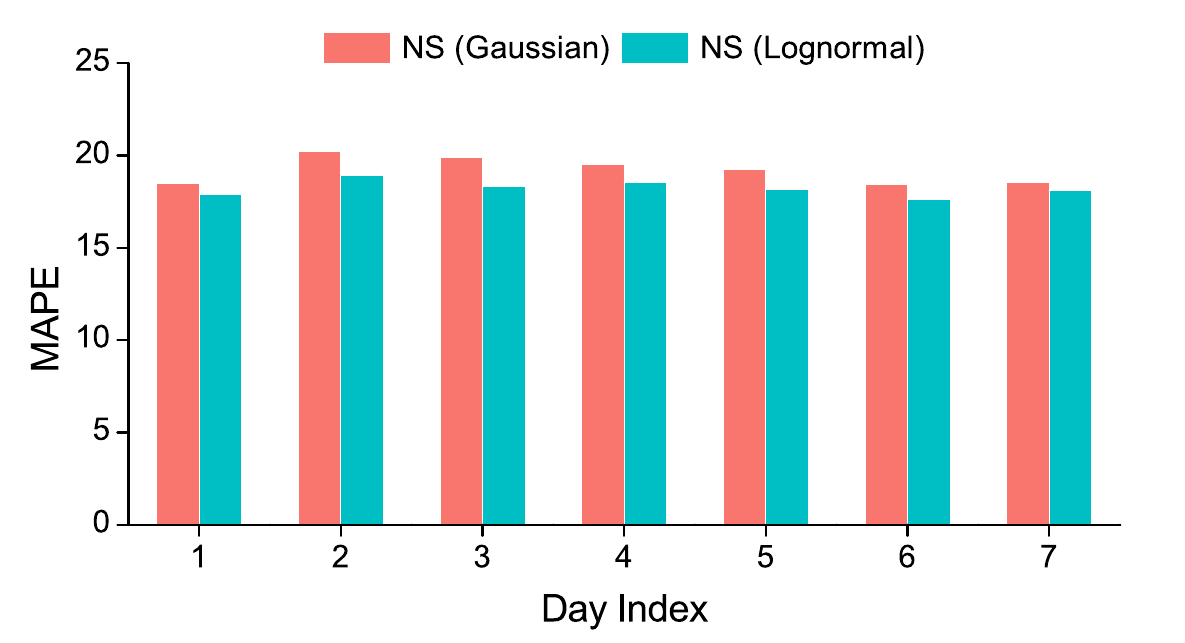}
		\caption{Non-Stationary method}
		\label{fig:log_NS}
	\end{subfigure}
	\caption{Performance comparison of the proposed methods under Gaussian and log-normal assumptions.}
	\label{fig:log}
\end{figure}

{\bf Exploiting Log-normality assumptions:} To begin with, the performance of the proposed methods under Gaussian and log-normal assumptions were analyzed to check the validity of log-normal properties of the data by incorporating the same in prediction scheme. Figure~\ref{fig:log} show the MAPE values obtained when the proposed methods were implemented under Gaussian and log-normal assumption. Figures clearly show that the predictions are better under log-normal assumption compared to Gaussian. Hence, it can be concluded that incorporating log-normal nature of the travel time data in to the modeling process did improve the prediction accuracy of the proposed methods and hence for further evaluation, the performance of the methods was analyzed only under log-normal domain.

\begin{figure}[h!]
 	\centering
 	\includegraphics[width=\linewidth]{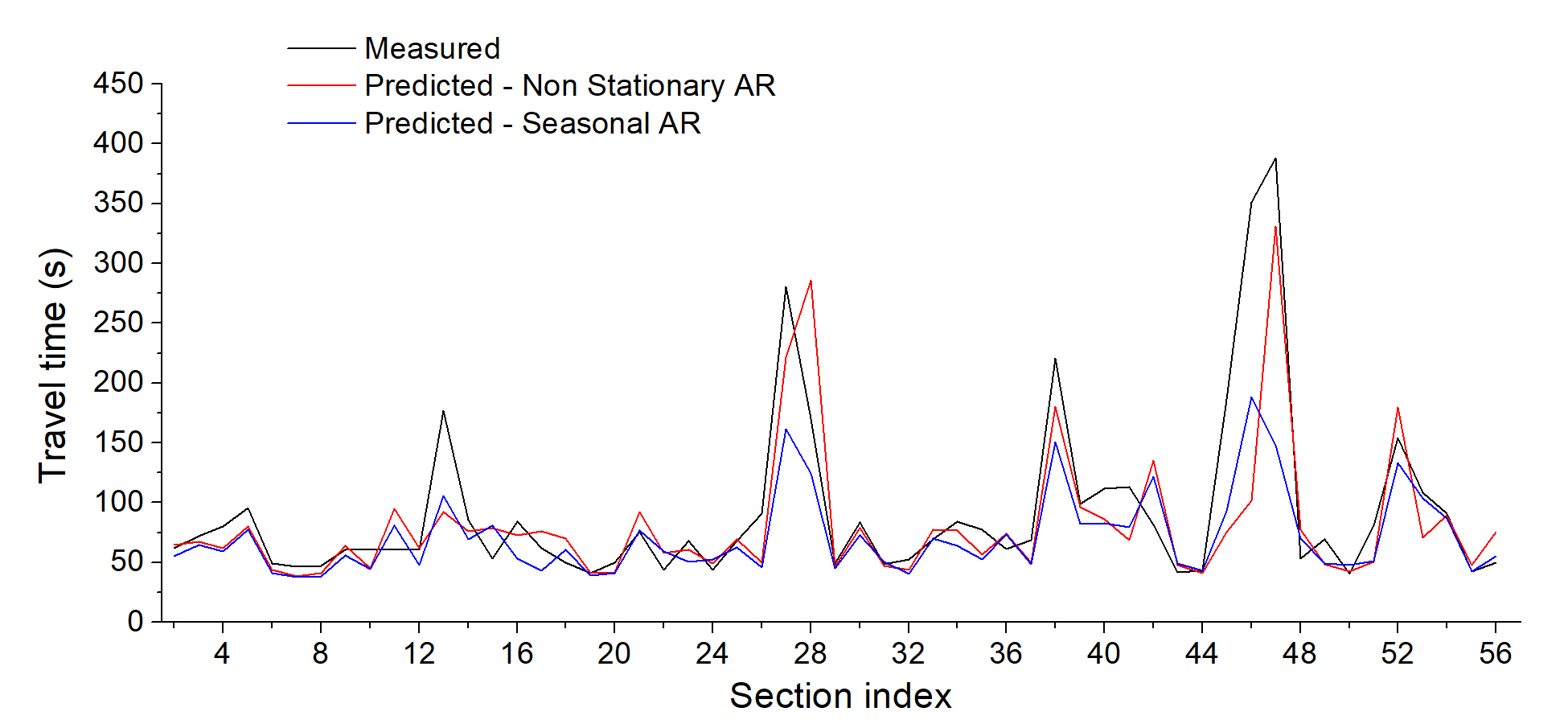}
 	\caption{Predicted and measured travel times for a sample trip.}
 	\label{fig:actpred}
 \end{figure}

In the next stage, performance was compared at an aggregate level of the seasonal AR and non-stationary method with the measured travel times from field. Sample plot of actual and predicted travel time values for a given trip (during test period) are shown in Figure~\ref{fig:actpred}. Figure clearly shows that the predicted travel time values are closely matching with the actual travel time values. Though, the performance of both the proposed methods are comparable, it can be observed that the peak travel time values around are better captured by non-stationary method.

\FloatBarrier

 \begin{figure}[!h]
	\centering
	\includegraphics[width=\linewidth]{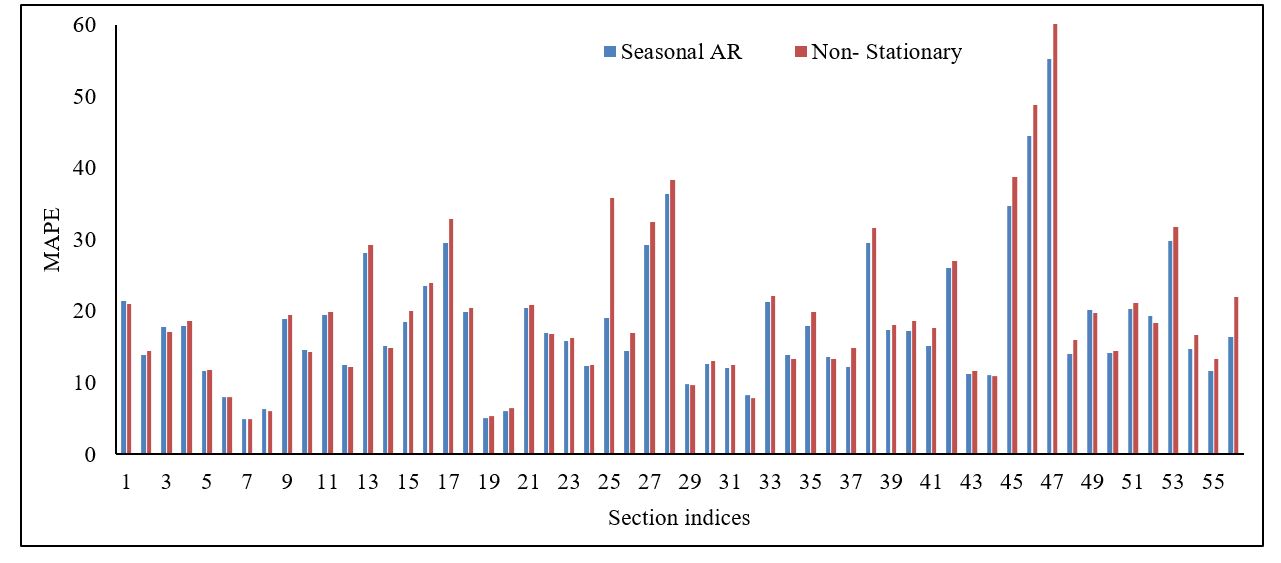}
	\caption{MAPE values obtained for all sections across the route.}
	\label{fig:sectionlevel}
\end{figure}
\begin{figure}[!h]
	\centering
	\includegraphics[width=\linewidth]{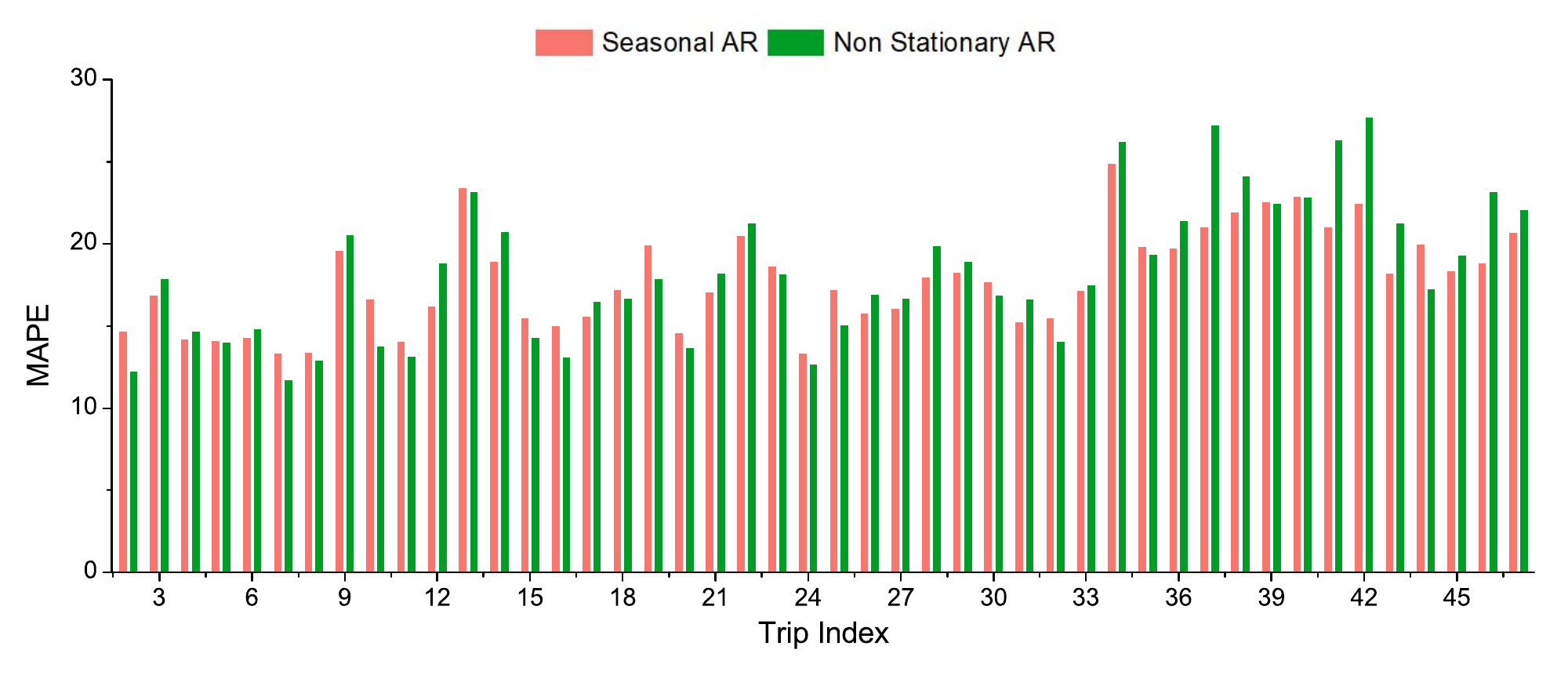}
	\caption{MAPE values obtained for all trips on a sample day.}
	\label{fig:triplevel}
\end{figure}

In the next level, comparisons at different levels like section level, and trip level are explored. For section level comparison, MAPE of predicted values of all trips for every section were aggregated and compared between the proposed methods. Figure ~\ref{fig:sectionlevel} shows the MAPE values for all trips for every sections. For trip level comparison, MAPE of predicted travel time values were aggregated over sections for every trip that happened on a sample day. Figure~\ref{fig:triplevel} shows the MAPE values obtained for all trips on a sample day. From Figure ~\ref{fig:sectionlevel} and~\ref{fig:triplevel}, it can be observed  that the performance of both the proposed methods are comparable with a slight upper-hand in accuracy by  seasonal AR.

\subsubsection{Multi-step prediction}The performance of the proposed methods while predicting multiple sections ahead was evaluated next, as a complete BATP system is expected to
give prediction updates to the passengers waiting at the bus stop. The proposed methodologies learnt separate temporal models at each section and integrating these various learnt
models for performing (the final) spatial multi-step prediction accurately. As the users experience the errors in terms of actual deviations, the evaluation of the proposed methods were checked in terms of MAE, i.e., absolute deviation of the predicted travel time from the actual travel time. 

\begin{figure}[h]
  	\centering
  	\begin{subfigure}{0.9\textwidth}
  		\centering
  		\includegraphics[width=0.9\linewidth]{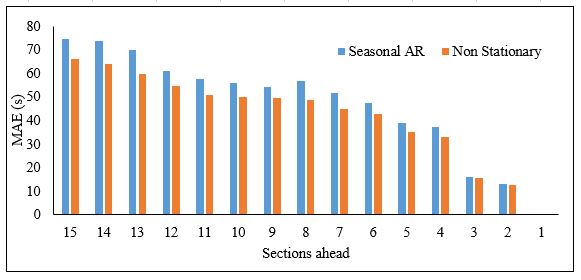}
  		\caption{Navallur (7.52 km from origin)}
  		\label{fig:navalur}
  	\end{subfigure}
  	\begin{subfigure}{0.9\textwidth}
  		\centering
  		\includegraphics[width=0.9\linewidth]{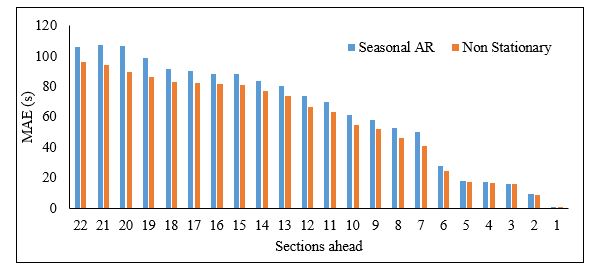}
  		\caption{Semmancheri (11.03 km from origin)}
  		\label{fig:semmancheri}
  	\end{subfigure}
  	\caption{Errors obtained from various multi-step predictions for selected bus stops.}
  	\label{fig:multistep}
  \end{figure}

The multi-step prediction algorithm predicts the travel time in all upcoming sections (56-$m$) from the current section ($m$) i.e. if the bus is at $m$\textsuperscript{th} section,
the multi-step prediction algorithm can predict travel time for all remaining sections till destination. The predicted values of all the upcoming sections were used to calculate the
Estimated Arrival Time (ETA) of a bus to any given bus stop that is located at section $n$. In particular, ($n$-$1$) times ETA can be calculated and predictions can be updated for
any bus stop located at section $n$. Figure ~\ref{fig:navalur} shows the comparison of MAE values obtained by predicting the arrival times at Navallur bus stop for its current
positions at every preceding section. This bus stop is located at 7.52 km from the origin, which is at $16$\textsuperscript{th} section. Hence, the arrival time predictions were
made for 15 different current positions of the bus starting from section $1$ to section $15$. Similarly, $22$ predictions were made for Semmencheri bus stop   that is located at
11.03 km from origin and a comparison of MAE values are presented in Figure ~\ref{fig:semmancheri}. {\em From the Figure ~\ref{fig:multistep}, it can be seen that the proposed
non-stationary method is performing better than the seasonal AR in all the cases.} 
 Overall, the above shown comparisons gives fair insights about the contributions of the proposed work.
  
\subsection{Performance Comparison}
\begin{figure}[htbp]
	\centering
	\includegraphics[width=\linewidth]{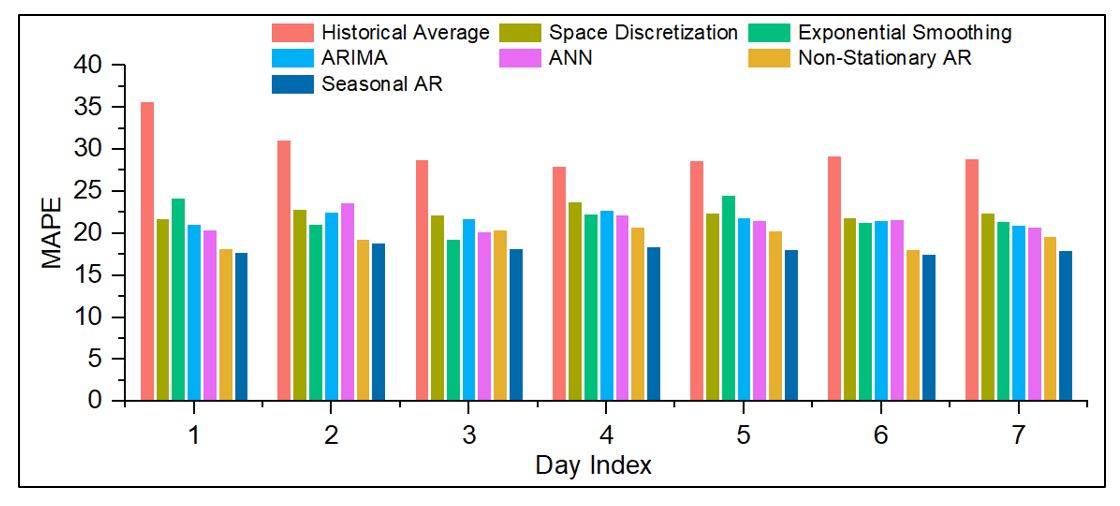}
	\caption{Comparison of MAPE values of various approaches for all days during test period.}
	\label{fig:allmethods}
\end{figure}

\noindent Further, the performance of the proposed prediction method was compared with space discretization \cite{leli:09}, time series approaches: exponential smoothing \cite{anil:18} and ARIMA \cite{jairam:18}, machine learning approach: ANN \cite{vivek:17} and historical average methods. To keep them all comparable with the proposed method, 27 days data were used for model building and one-week data to test the performance. In the space discretization approach \cite{leli:09}, travel time of a bus in an upcoming section was predicted using the travel time in the previous section alone i.e. the model hypothesized a relation in travel time between neighbouring subsections. As part of implementation, the space discretization approach uses the travel time data obtained from previous two buses to predict the travel time of the next bus. An ANN based data-driven approach \cite{vivek:17} was trained with four independent variables i.e. previous four significant trips travel time on the same segment, to predict the next bus travel time. For training the neural network model, a multi-layer feed-forward network with the Leveberg-Marquardt back propagation algorithm was used. Exponential smoothing predicts the travel time of the next bus using the most recent observation and the most recent estimate from the previous bus. Another time series approach, ARIMA predicted route travel time and then segment travel times using segment-specific coefficients, which were established using historic database \cite{jairam:18}. The errors obtained from all these methods are presented in terms of MAPE. Figure~\ref{fig:allmethods} shows the comparison of MAPE values obtained by aggregating all trips at all section at day level for all the methods.

\FloatBarrier

From Figure~\ref{fig:allmethods}, it can be observed that both the proposed methods outperforms existing methods on all test days. Amongst the proposed methods, seasonal AR performance can be found to be slightly better than non-stationary method. Such marginally lower performance of the NS method can be attributed to the lack of sufficient quality data. In this paper, we used 27 days  data for training. The non-stationary model essentially tries to learn a linear regressor at every time bin with 27 data points. Hence, it can be expected that the performance in this case will improve with more quality data. However, the classical approach learns only one common linear regressor at every section and hence more data may not affect the performance. Thus, it can be concluded that the proposed methods in this study are viable for the prediction of bus arrival times under mixed traffic condition. 

\section{SUMMARY AND CONCLUSIONS}

Providing information about bus arrival times/travel times at bus stops helps in reducing the uncertainties and waiting time of commuters. Bus travel times are often affected by
diverse factors such as weather, incidents, signals, bus stops, bottlenecks etc., making the bus arrival time prediction complex in an urban traffic environment. The present work
developed a bus arrival time prediction exploiting the temporal correlations of the AVL data. The unique aspects and main contributions of the study include:
\begin{itemize}
\item The choice of observations in modelling the time series is very different from existing approaches. The travel times experienced at different times of the day at a particular 
section constitute the observations of our time series. This choice of observations is motivated from the necessity to learn the temporal correlations in the AVL data.
\item We explored the SAR time series approach using both multiplicative and additive seasonal model and finally chose the best among them (based on the AIC criteria) at each section.
\item  We  proposed a novel sequential non-stationary model for learning temporal correlations. 
This model and its associated learning is not standard from a statistics OR  time series analysis perspective and we believe is a 
novel  contribution.
\item Further, utilizing the above two class of sequential models (learnt at each section) for solving the final multi-section ahead bus travel time prediction is not straightforward. 
We discussed the issues involved and proposed a prediction framework to  achieve this, which makes our method viable for real-time implementation. This framework is  general enough to 
be used in conjunction with any of the proposed temporal prediction models.
\item Lastly, we exploit the log-normal nature of the travel time distributions into the predictions to obtain statistically optimal predictions under log-normal assumptions. 
Our experiments on real data clearly vindicate this choice by revealing superior predictions (under log-normal assumptions) in comparison with Gaussian assumptions on the predictive model.
\end{itemize}
Our experiments clearly demonstrated the superior performance of the proposed methods in comparison with the existing approaches.  Hence, this study concludes that the proposed methods are viable to  implement for a bus arrival time prediction system under heterogeneous traffic condition  with a good data base availability.

A natural future work would be to explore non-linear models to capture the temporal correlations and then use our proposed multi-step prediction framework for the final ETA
prediction. In this work we  fully exploited only the temporal correlations in the AVL data. Another potential direction for future research would be to exploit the more general spatio-temporal
correlations in the data using space-time ARIMA  models.  
\comment{
To start with, travel times were analyzed to assess its characteristics. A goodness of fit test applied on the hourly section wise travel times revealed a predominantly log-normal
behaviour in the marginal distributions. The existing methods to tackle the travel time prediction problem were  assuming a Gaussian distribution to the data.  In addition, majority
of them were partly oblivious to the temporal correlations hidden in the data. The above two key observations lead us to modeling the data as a time series under a log-normal
distributional assumption. Under two broad approaches to tackle the non-stationarity of the data, model fitting methods and their associated statistically optimal prediction schemes
were proposed. To the best of our knowledge, the model and its associated learning under non-stationary method is neither conventional nor standard from a statistics or time series
analysis perspective and is a novel contribution. The proposed methodologies were found to be performing better than all the considered existing approaches under log-normal
assumptions. In addition, a real-time  multi-segment ahead (spatial) prediction framework is proposed which is general enough to be used in conjunction with any of the proposed
temporal prediction models. Hence, this study concludes that the proposed methods are viable to  implement for a bus arrival time prediction system under heterogeneous traffic
condition  with a good data base availability. }

\section{ACKNOWLEDGMENTS}
\noindent The authors acknowledge the support for this study as a part of the project RB/16-17/CIE/001/ TATC/LELI under the Development of a Dynamic Traffic Congestion Prediction System for Indian Cities, funded by Tata Consultancy Services.

\bibliographystyle{unsrt}


\end{document}